%% file: main.tex
\documentclass[conference]{IEEEtran}

\newif\iftechreport
\techreporttrue
\iftechreport
  \newcommand{\omitproofdisclaimer}[1]{}
  \newcommand{\insertproof}[1]{#1}
  \newcommand{\insertFullSection}[1]{#1}
  \newcommand{\insertShortSection}[1]{}
  \newcommand{\insertproofsketch}[1]{}
  
\else
  \newcommand{\omitproofdisclaimer}[1]{#1}
  \newcommand{\insertproof}[1]{}
  \newcommand{\insertFullSection}[1]{}
  \newcommand{\insertShortSection}[1]{#1}
  \newcommand{\insertproofsketch}[1]{#1}
  
\fi

\usepackage[textwidth=1.5in,shadow,backgroundcolor=yellow,textsize=tiny]{todonotes}

\newcommand{\itamar}[1]{\todo[color=green!20]{Itamar: #1}}

\usepackage{xcolor}
\usepackage[compress]{cite}
\usepackage{microtype} 
\usepackage{amsmath}
\usepackage{amsthm}
\usepackage{tabto}
\usepackage{algpseudocode}
\usepackage{xspace}
\usepackage{paralist}
\usepackage{url}
\PassOptionsToPackage{hyphens}{url}
\usepackage{subfloat}
\usepackage[caption=false]{subfig}  
\usepackage[utf8]{inputenc}
\usepackage{wesa}
\usepackage[T1]{fontenc}

\usepackage{amssymb}
\usepackage{graphicx}
\usepackage{tabularx}
\usepackage{array}
\usepackage{hhline}
\usepackage{multirow}
\usepackage{epsf}
\usepackage{epstopdf}
\usepackage{algpseudocode}
\usepackage[Algorithm]{algorithm}
\usepackage{color, colortbl}
\usepackage{subfiles}

\usepackage{tikz}
\usetikzlibrary{automata,positioning}
\usepackage{tikz-qtree}

\usepackage{siunitx}
\usepackage{verbatimbox}
\usepackage{pgfplots}
\pgfplotsset{width=7cm,compat=1.3}
\usepgfplotslibrary{units}

\usepackage{footnote}
\makesavenoteenv{tabular}
\makesavenoteenv{table}

\newcommand{\ignore}[1]{}

\DeclareRobustCommand{\inlinelist}[1]{\begin{inparaenum}[(i)] #1 \end{inparaenum}}

\makeatletter
\def\blfootnote{\xdef\@thefnmark{}\@footnotetext}
\makeatother

\newtheorem{theorem}{Theorem}
\newtheorem{corollary}[theorem]{Corollary}

\newtheorem{proposition}[theorem]{Proposition}

\algnewcommand{\AND}{\textbf{and}\xspace}
\algnewcommand{\OR}{\textbf{or}\xspace}

\newcommand{\TableFontSize}{\footnotesize}

\newcommand{\abs}[1]{\left\vert#1\right\vert}
\newcommand{\set}[1]{\left\{#1\right\}}

\newcommand{\eps}{\varepsilon}

\newcommand{\dr}{\eps} % Decline ratio
\newcommand{\nbins}{n}   % Number of rsrcs
\newcommand{\nfreebins}{k} % (minimal) number of free bins
\newcommand{\nsched}{s}  % number of schedulers (allocators)
\newcommand{\nhappy}{H_{\nsched}} % (expected) number of happy schedulers, namely, schedulers which successfully allocates a VM
\newcommand{\npothappy}{F_{\nsched}} % An rv, representing the # of schedulers which sample at least one free bin
\newcommand{\hk}{H^{k}_{\nsched}} % rv for number of happy agents when there're k free bins
\newcommand{\hkpp}{H^{k+1}_{\nsched}} % rv for number of happy agents when there're k+1 free bins
\newcommand{\phk}{F^{k}_{\nsched}} % rv for number of potentially happy agents when there're k free bins
\newcommand{\phkpp}{F^{k+1}_{\nsched}} % rv for number of potentially happy agents when there're k+1 free bins
\newcommand{\dif}{D} % dif (k, f) = E \left[\hk | \phk = f+1\right] - E [\hk | \phk = f]
\newcommand{\PrFF}{\sigma} % Prob' for a sched' to Find a Free (i.e., not totally full) bin.
\newcommand{\budget}{B} %sd budget
\newcommand{\estkalg}{EstimateK}
\newcommand{\maxparalg}{MaximizeParallelism} %Alg' to maximize parallelism in the balls-and-bins model
\newcommand{\algtop}{APSR}%Alg' to generate / destroy sched's in a real system
\newcommand{\algtopKavg}{\algtop$_{avg}$}
\newcommand{\rand}{Rand}
\newcommand{\host}{\vec{h}} 
\newcommand{\hosts}{\mathbf{H}} 
\newcommand{\request}{\vec{r}}
\newcommand{\requests}{\mathbf{R}}
\newcommand{\flavors}{\mathbf{C}}
\newcommand{\flavor}{\vec{c}}
\newcommand{\dqueries}{d}

\begin{document}
\title{Parallel Virtual Machines Placement with Provable Guarantees
% \title{Parallel Virtual Machines Deployment with Provable Guarantees
% \thanks{ISBN 978-3-903176-39-3© 2021 IFIP}
}

\author{
\IEEEauthorblockN{Itamar Cohen}
\IEEEauthorblockA{
% \textit{Department of Electronics and Telecommunications} \\
\textit{Politecnico di Torino}\\
Torino, Italy \\
itamar.cohen@polito.it}
\thanks{This work was done while the first author was with Ben-Gurion University.}
\\
\IEEEauthorblockN{Yaniv Sa'ar}
\IEEEauthorblockA{\textit{Independent} \\
Israel \\
yaniv.saar.mail@gmail.com}
\and
\IEEEauthorblockN{Gil Einziger}
\IEEEauthorblockA{
\textit{ %Department of Computer Science, 
Ben-Gurion University of the Negev} \\
Beer Sheva, Israel \\
gilein@bgu.ac.il}
\\
\IEEEauthorblockN{Gabriel Scalosub}
\IEEEauthorblockA{
\textit{%School of Electrical and Computer Engineering, 
Ben-Gurion University of the Negev} \\
Beer Sheva, Israel \\
sgabriel@bgu.ac.il}
\and
\IEEEauthorblockN{Maayan Goldstein}
\IEEEauthorblockA{\textit{Independent} \\
Israel \\
maayan.goldstein@gmail.com}
\\
\IEEEauthorblockN{Erez Waisbard}
%\IEEEauthorblockA{\textit{Independent} \\
\IEEEauthorblockA{\textit{The Open University}\\
Ra’anana, Israel \\
%erezcrypt@gmail.com}
erezwa@openu.ac.il}
}

\pagestyle{plain}

\IEEEoverridecommandlockouts
% \IEEEpubid{
% \makebox[\columnwidth]{
% ISBN 978-3-903176-39-3~\copyright~2021 IFIP \hfill
% }
% \hspace{\columnsep}
% \makebox[\columnwidth]{ }
% }

\maketitle

\pagestyle{empty}

% \pubid{ISBN 978-3-903176-39-3~\copyright2021 IFIP}
\IEEEpubidadjcol

\begin{abstract}
Network Function Virtualization (NFV) carries the potential for on-demand deployment of network algorithms in virtual machines (VMs). 
In large clouds, however, VM resource allocation incurs delays that hinder the dynamic scaling of such NFV deployment.
Parallel resource management is a promising direction for boosting performance, but it may significantly increase the communication overhead and the decline ratio of deployment attempts.
Our work analyzes the performance of various placement algorithms and provides empirical evidence that state-of-the-art parallel resource management dramatically increases the decline ratio of deterministic algorithms but hardly affects randomized algorithms.
We, therefore, introduce APSR -- an efficient parallel random resource management algorithm that requires information only from a small number of hosts and dynamically adjusts the degree of parallelism to provide provable decline ratio guarantees. 
We formally analyze APSR, evaluate it on real workloads, and integrate it into the popular OpenStack cloud management platform. Our evaluation shows that APSR matches the throughput provided by other parallel schedulers, while achieving up to 13x
lower decline ratio and a reduction of over 85\% in communication overheads. 
\end{abstract}

\setlength\tabcolsep{5 pt}
\section{Introduction}
\label{sec:intro}

\blfootnote{An earlier version of this work was published in~\cite{APSR_IFIP}.}
% \blfootnote{* The work was done while this author was with Ben-Gurion University.}
The {\em Network Function Virtualization (NFV)} paradigm enables network infrastructure to be virtually deployed on standard cloud infrastructure.
Specifically, NFV allows running firewalls, deep packet inspection, load balancing, and monitoring without relying on physical middleboxes~\cite{Middleboxes,EASE}.
NFV is composed out of (often long) service chains that each packet needs to traverse. One of the main advantages of NFV is the ability to scale the service chain on demand without any physical change to the network. 
%One of the key arguments for NFV is that it enables online deployment of network services, as well as allowing scaling of such services according to the current workload requirements.
%These features should presumably improve the overall system performance in various perspectives, including throughput and latency.
Unfortunately, current cloud placement is not optimized for high-throughput placement, making large service chains slow to deploy. 

%these improvements are not manifested in large clouds, as we see and discuss in the sequel.

In principle, once the user issues a request to allocate a new {\em Virtual Machine (VM)}, a {\em scheduler} selects a host to accommodate the VM. 
While the deployment time of optimized VMs or containers (e.g., using Kubernetes) can be tens of milliseconds~\cite{ClickOS}, selecting a host on which to place the VM may require hundreds of milliseconds in large clouds~\cite{OSProblem,KubeProblem,ASC}.
It follows that the potential performance boost of using NFV remains largely unfulfilled in large clouds due to bottlenecks in scheduling deployment requests.
% While few works optimize decision times, other aspects of VM deployment are drastically optimized  For example, VMs (and containers) can be deployed within tens of milliseconds~\cite{ClickOS}.
%It follows that the potential performance boost of using NFV remains largely unfulfilled in large clouds due to bottlenecks in scheduling deployment requests.

The main reason that the host selection process takes so long is that most current 
resource management algorithms~\cite{Power, Power2, Yao2013, FaultTolerant, Switching,Yaniv2,RazPlacement,SHABEERA2017616}  require complete information about the availability of resources on the system's hosts.
In a large cloud, gathering the current state from hundreds and sometimes thousands of hosts translates to high communication overheads, resulting in a performance bottleneck~\cite{ASC, KubeProblem}.
In particular, some experiments show that when the number of hosts is above 400, more than 90\% of the scheduling time is wasted in collecting the fresh system's state information~\cite{OSProblem}. 

Intuitively, one could boost throughput by running multiple schedulers in parallel. However, such an approach may translate to multiple schedulers trying to place requests simultaneously on the same host, leading to race scenarios~\cite{Omega, Sparrow, host_subset_size}. 
In such cases, not all the requests will be successful, and the host may {\em decline} some of the requests.
This decline translates to having the scheduler retry
to serve the same request, resulting in excessive latency.
This added latency may be unacceptable in an NFV environment, which may have to respond to bursts of requests, e.g., when the system must respond to a flash crowd or a cyber attack~\cite{ASC}.
Hence, a provider is typically required to bind the {\em decline ratio}, namely, the ratio between the number of declined requests and the total number of requests. The maximum allowed decline ratio is typically defined in the {\em Service Level Agreement (SLA)}~\cite{NFV_RA_survey16, ASC}, or in the {\em Key Performance Indicators (KPIs)} ~\cite{okpi}.

An efficient VM placement algorithm should, therefore, strive to
\begin{inparaenum}[(i)]
\item increase parallelism, while
\item maintaining a low communication overhead, and 
\item ensuring a bounded decline ratio.
\end{inparaenum} 
However, to the best of our knowledge, no previous work has studied the interplay between these conflicting aspects.

{\bf Our contributions.}
Our work starts by studying the impact of parallelism on the decline ratio of various popular placement algorithms. We show that parallelism may drastically increase the decline ratio, where we attribute this increase to the determinism of most algorithms. Interestingly, we find that randomly placing VMs in suitable hosts allows for a large degree of parallelism without a significant impact on the decline ratio. That is, random placement is very efficient in parallel settings.  Our study further shows that in the random policy, the decline ratio depends on the number of parallel schedulers and the number of hosts that can accommodate each VM. 
In general, low-utilization environments allow for more schedulers than high-utilization ones. 

Equipped with these observations, we introduce our proposed algorithm, \algtop, that dynamically adjusts the number of parallel schedulers according to the system's utilization and incorporates randomness into its decision making.
\algtop\ guarantees that the expected decline ratio is always within
a predefined requirement. Furthermore, \algtop\ is inherently optimized to query only a small number of hosts, thus reducing the communication overheads. %and allows for higher throughput. 

We formally analyze the performance of \algtop\ where we provide guarantees as to its communication overhead and its expected decline ratio.
We also evaluate the performance of \algtop\ for three real-life datasets and show that it enables a high degree of parallelism (e.g., effectively running 20-100 schedulers) in various realistic scenarios. We further show that \algtop\ reduces the communication overhead by over 85\% compared to state-of-the-art algorithms. Finally, we integrate and implement \algtop\ within the OpenStack framework and show that it matches the throughput of the fastest OpenStack configuration while significantly reducing the decline ratio and the communication overhead.%

\section {Related Work}
\label{sec:background}

This section provides a short survey of commonly used VM placement paradigms. For each such approach, we discuss the various algorithms that apply it in their design. We further provide insight into the main differences between our suggested solution and these algorithms, summarized in Table~\ref{tbl:taxonomy}.

\begin{table*}[t]
	\centering
\caption[Comparison of approaches for scheduling requests in a multi-host system]{Comparison of approaches for scheduling requests in a multi-host system. The different approaches are compared in terms of their
\inlinelist{
\item throughput (rate of assignment attempts),
\item decline ratio (expected ratio of attempts that fail), and
\item overhead (amount of communication/synchronization required to gather the state information for making an assignment decision).
}
For each of the approaches, we provide some concrete examples of existing architectures that implement the approach.}
\label{tbl:taxonomy}
	\renewcommand\arraystretch{1.2}
    \TableFontSize
    \begin{tabular}{|p{1.2 cm}|c|c||c|c|c||c|}
		\hline
		Approach & \#Schedulers & Description & Throughput & Decline Ratio & Overhead & Examples \tabularnewline %& Example
		\hline \hline
		{\multirow{4}{0pt} {Global Snapshot}}
		& 
		{\multirow{2}{*}{Single}} &
		Monolithic & Low & Low, guaranteed & High & Maui~\cite{Maui14} \\ 
        \cline{3-7}
 		& & Cached Snapshot & Low & Low & Low & ASC~\cite{ASC} \\ 
        \cline{2-7}
 		&
 		{\multirow{2}{*}{Fixed}} & Multiple Snapshots & High & High & High & OpenStack~\cite{filterscheduler}\\ 
        \cline{3-7}
 		& & Shared Snapshot & High & High & Low-Mid & Omega~\cite{Omega} \\ 
		\hline \hline
		{\multirow{2}{0pt} {Partitioning}}
 		&
 		{\multirow{2}{*}{Fixed}} & Static Partition & Mid & Mid & Low & Quincy~\cite{Quincy} \tabularnewline 
        \cline{3-7}
 		& & Dynamic Partition & Mid & Mid & Low-Mid & Mesos~\cite{Mesos} \tabularnewline 
		\hline \hline
 		Sampling & Adaptive & Adaptive sample size & Mid-High~\footnotemark & Low, guaranteed & Low, guaranteed & Our \algtop\ algorithm \tabularnewline 
        \hline
	\end{tabular}
\end{table*}

{\bf The global snapshot-based approach.}
Traditionally, placement algorithms take a snapshot of the entire system's state before handling each request.
This precise state information allows for a single {\em monolithic} scheduler 
(e.g., Maui~\cite{Maui14}, and the single-scheduler algorithms proposed in~\cite{Javad_APSR}) to select a host to accommodate the request while prioritizing the hosts in some manner. 
The monolithic approach guarantees a low decline ratio, as the scheduler
operates alone on an up-to-date view of the available resources.
However, the per-request overhead of this approach is substantial due to both the communication overhead of querying all hosts~\cite{OSProblem,ASC, KubeProblem} and the latency of computing the placement decision itself, which may take several seconds~\cite{Omega}.
Such a long latency might be reasonable when scheduling large batch jobs (e.g., in HPC environments) but is prohibitively costly when a prompt reaction is critical, e.g., scaling out a service chain's capacity due to an increase in demand. 

One of the ways suggested for decreasing the overhead of the monolithic scheduler is to periodically {\em cache} a snapshot of the system's state~\cite{ASC}. 
However, when the cached state becomes stale, the scheduler may be unaware of resources that have recently become available, resulting in an increased number of needlessly declined requests.
Furthermore, this approach achieves low throughput as it only employs a single scheduler.

Running multiple schedulers in parallel is a straightforward technique to increase throughput. Indeed, OpenStack allows for multiple parallel schedulers to increase the throughput~\cite{filterscheduler}. 
However, our work shows that running multiple independent schedulers translates to collisions when multiple schedulers select the same hosts simultaneously. Such collisions result in excessive decline ratios, degrading performance. Interestingly, the OpenStack community acknowledges this problem and mitigates its impact by allowing the user to add a certain degree of randomness to the schedulers~\cite{host_subset_size, Sparrow}. 
Further, the seminal work of~\cite{Omega} shows that when system utilization is high, Google's schedulers require more than two attempts to place each request.
Our work shows (in Section~\ref{sec:PlacementStudy}) that parallel scheduling yields high decline ratios for a variety of placement algorithms and that random placement is more robust than deterministic placement. Intuitively, deterministic algorithms select the same "best" host, rendering them inferior to random algorithms. 

To insert some degree of randomness into the scheduling process, the OpenStack community introduced the parameter {\tt scheduler\_host\_subset\_size}~\cite{host_subset_size} (denoted $\Lambda$), which works as follows:
After ranking the available hosts,
the scheduler randomly assigns the request to one of the top-$\Lambda$ ranking hosts. 
In the absence of a rigorous theory studying the effect of $\Lambda$ on the system's performance, its value is commonly determined using crude estimations and rules of thumb. 
Our work helps to configure the parameter $\Lambda$ properly.
Furthermore, in Section~\ref{sec:PlacementStudy}, we show that the common approach of setting $\Lambda$ as a small constant results in poor performance. 

The Omega scheduler~\cite{Omega} suggests a new approach that aims at optimizing the usage of a global snapshot by multiple schedulers. Omega decreases the communication overheads by allowing multiple schedulers to share the state information. However, Omega does not provide guarantees on the decline ratio. Thus, our approach is also useful in Omega's framework. 

{\bf The partitioning-based approach.}
{\em Partitioning} the hosts between different schedulers is a simple approach that removes conflicts between schedulers and decreases the pre-placement communication overheads as each scheduler only acquires state information about some of the hosts. Quincy~\cite{Quincy} uses a {\em static} partition, which occasionally results in non-compulsory declines due to fragmentation of resources~\cite{Omega}. Namely, a scheduler may fail to place a request in its partition, even if hosts in other partitions can accommodate the request. 
Mesos~\cite{Mesos} suggests using {\em dynamic} partitioning, where a central controller dynamically allocates hosts to schedulers on demand to minimize fragmentation at the expense of complexity. 
Note that Quincy and Mesos provide no guarantees on the impact of fragmentation on the decline ratio. 
%Note that the partitioning-based approach provides no guarantees on the decline ratio.

{\bf The sampling-based approach.}
%
% \textbf{The sampling-based approach in other problems.}
The sampling-based approach was extensively studied in the context of balanced allocation problems~\cite{parallel_P2,P2, Sparrow, Tarcil}. These problems essentially assume an (infinite) buffer for pending requests in each host, and the goal is to allocate requests to hosts in a way that minimizes the maximum load on all hosts.
The celebrated power-of-two-choices algorithmic paradigm~\cite{P2, parallel_P2} shows that sampling only a few (e.g., two) hosts and selecting the least-loaded sampled host provides strong guarantees on the expected maximal load. 
Sparrow~\cite{Sparrow} and Tarcil~\cite{Tarcil} implement variants of this concept in concrete cloud environments.

However, balanced allocation problems are inherently different from the ones addressed in our work as they consider infinite capacity hosts that never decline requests, and instead, their algorithms make an effort to balance the load evenly~\cite{P2, parallel_P2, Sparrow, Tarcil}. In contrast, we consider finite-capacity hosts that decline requests that exceed their capacity limitations, making  load-balancing-based algorithms incomparable with our work. 
That said, our \algtop\ is part of the sampling-based approach as it queries a small number of hosts, and while load-balancing based algorithms provide guarantees on the maximum load~\cite{P2, parallel_P2, Sparrow, Tarcil}, \algtop\ provides guarantees on the decline ratio. 

\section{System Model for Parallel Scheduling}
\label{sec:model}

We consider a collection $\hosts$ of $\nbins$ hosts where each host has some multi-dimensional capacity corresponding to several types of resources, e.g., memory, CPU, or disk space. Formally, we model each $\host \in \hosts$ as a vector whose coordinates correspond to the currently available resources of each type. We refer to this vector as the {\em state} of the host.
We further consider a collection $\requests$ of \emph{requests}, each modeled as a vector of demand for each resource. 
We assume each request $\request \in \requests$ has its vector drawn from some finite set of possible request vectors, or {\em flavors}, $\flavors=\set{\flavor_1,\ldots,\flavor_m}$.
A host $\host$ is considered \emph{available} for request $\request$ if it has enough resources of each type, i.e., if 
$\request \leq \host$, coordinate-wise.

\footnotetext{Throughput is inversely proportional to system utilization, and adapts to the amount of resources available in the system.}

We assume time is slotted, such that in every time slot, some requests arrive at the system and are queued, pending assignment to hosts.
% \itamar{Added footnote below.}
% \footnote{We use time slots as an abstract model for the problem. In practice, schedulers need not be synchronized.}
We denote by $\nsched$ the number of {\em parallel schedulers} that may perform scheduling decisions simultaneously in any single time slot.
In each time slot $t$, given a queue consisting of some $q$ requests pending at $t$, each scheduler dequeues a request.  Schedulers may query (sample) the state of some subset of hosts, and assign the request to an available host (if they queried such a host). 
We note that when $\nsched>1$, multiple schedulers may concurrently assign their pending requests to the same host.

Any host $\host \in \hosts$ resolves concurrent requests assigned to $\host$ at the same time slot in some arbitrary order. The resolution of request $\request$ being assigned by some scheduler to host $\host$ \emph{fails} if the host is no longer available when it resolves $\request$, and is \emph{successful} otherwise. The host updates its available capacity upon a successful resolution by setting $\host = \host - \request$.  Requests live for some time, and the host regains the resources used by completed requests. If request $\request$ placed on host $\host$ is completed we update the resource state of the host by setting $\host = \host + \request$.
The above model implies that a request fails if either \begin{inparaenum}[(i)]
\item the scheduler does not find an available host, or
\item the chosen host is no longer available once it resolves the request.
\end{inparaenum}

In every time slot $t$, and for every request flavor $\flavor \in \flavors$, we let $\nfreebins^{(t)}_{\flavor}$ denote the number of hosts in $\hosts$ that are available for a request of flavor $\flavor$ at time $t$. We further let $\nfreebins^{(t)}$ denote an {\em estimate} of the number of hosts that may accommodate {\em any} request that may arrive at time $t$. We note that $\nfreebins^{(t)}$ may be a pessimistic estimate (e.g., by setting $\nfreebins^{(t)}=\min_{\flavor} \ \nfreebins^{(t)}_{\flavor}$), or it may incorporate some information about the workload distribution, or otherwise the system state.
We will usually be omitting the superscript of $(t)$, and refer to $\nfreebins_{\flavor}$, and $\nfreebins$, when the time slot in question is clear from the context.

The {\em decline ratio} is the ratio between the number of failed assignment attempts and the total number of assignment attempts performed by the system.
We use $ \delta $ to denote the system's expected decline ratio (for some set of requests $\requests$). Since we are handling requests independently, $\delta$ is the a posteriori probability of having a declined assignment attempt.

We assume the system is subject to a {\em Service Level Agreement (SLA)} which requires that the decline ratio is at most $\dr$, for some $\dr \in [0,1]$.\footnote{Current algorithms are oblivious to such constraints, and might violate this requirement.
Our \algtop\ algorithm takes such constraints into account, and produces solutions that provably satisfy them.}
To control the overheads, we limit the maximal number of hosts queried (by all schedulers) in each time slot to $B$. 
In every time slot $t$, we denote by $\dqueries$ the number of hosts queried by any scheduler with a pending request at $t$. A \emph{valid configuration} of schedulers determines $\nsched$ and $\dqueries$, such that $\nsched \cdot \dqueries \leq \budget$, and the probability of a failed assignment attempt is at most $\dr$. We seek the valid configuration maximizing the number of parallel schedulers ($\nsched$).
% We refer to this problem as the {\em Constrained Maximum Parallelism (CMP)} Problem.
Table~\ref{tbl:notations_balls_n_bins} summarizes the notation used in our model, as well as further notation defined in later sections.

\begin{table}[t]
    \centering
    \caption{List of Symbols. The top section corresponds to our system model (Section~\ref{sec:model}), the middle section corresponds to our performance guarantees (Section~\ref{sec:comb_analysis}), and the bottom section corresponds to our evaluation (Section~\ref{sec:sim_algtop})}
    \label{tbl:notations_balls_n_bins}
    \TableFontSize
    \begin{tabular}{|c|p{5.8cm}|}
		\hline
		Symbol & Meaning \tabularnewline
		\hline
		\hline
		$\hosts$ & Set of hosts\tabularnewline
		\hline
		$\nbins$ & Number of hosts (bins) \tabularnewline
		\hline
		$\host$ & Host in $\hosts$ (resources availability vector)\tabularnewline
		\hline
		$\requests$ & Set of requests \tabularnewline
		\hline
		$\request$ & Request in $\requests$ (resources demand vector)\tabularnewline
		\hline
		$\flavors$ & Set of requests flavors\tabularnewline
		\hline
		$\flavor$ & Flavor in $\flavors$ of a request \tabularnewline
		\hline
        $\nsched$ & Number of schedulers (agents)\tabularnewline
		\hline
        $\delta$ & Actual decline ratio\tabularnewline
		\hline
        $\dr$ & Maximum allowed decline ratio by the SLA\tabularnewline
		\hline
		$\budget$ & Budget for overall number of queries\tabularnewline 
		\hline
		$\dqueries$ &Number of hosts queried by each scheduler \tabularnewline
		\hline
        $\nbins_{\flavor}$ & Number of hosts queried for requests of flavor $\flavor$\tabularnewline
		\hline
        $\nfreebins_{\flavor}$ & Number of available hosts for flavor $\flavor$\tabularnewline
		\hline
        $\nfreebins$ & Number of available hosts for any request \tabularnewline
		\hline
		\hline
        $\npothappy$ & Number of potentially happy agents\tabularnewline
		\hline
        $\nhappy$ & Number of happy agents\tabularnewline
		\hline
        $\PrFF$ & See~\eqref{Eq:def_sigma} 
        \tabularnewline
		\hline
		$Bin(a,b,c)$ & See~\eqref{Eq:def_bin} 
        \tabularnewline
		\hline
		\hline
		$\lambda_a$ & Poisson arrival rate \tabularnewline
		\hline
		$\lambda_d$ & Poisson departure rate\tabularnewline
		\hline
	\end{tabular}
	\normalsize
% 	\end{tabularx}
\end{table}

\section{Parallelism and Placement Algorithms}
\label{sec:PlacementStudy}
We begin by evaluating the effect of parallel schedulers on the decline ratio of existing placement algorithms.

\subsection{Evaluated Algorithms}
\label{sec:evaluated_algorithms}
We briefly introduce some common placement algorithms. For further details, see, e.g., ~\cite{Mills11comparingvmplacement}. 

OpenStack's default placement algorithm is the {\em WorstFit (WF)} algorithm~\cite{filterscheduler}. WF places requests on one of the least loaded hosts to maximize the hosts' remaining resources. For the multi-dimensional settings, we implement a pessimistic variant of WF, where we consider a host load to be the maximum load over all the possible resources.

The \emph{FirstFit (FF)}~\cite{Epstein2003}
algorithm assigns a request to the first available host, assuming some arbitrarily fixed ordering of the hosts. This approach aims at minimizing the number of utilized hosts, thus reducing energy consumption.

The \emph{Adaptive} algorithm~\cite{RazPlacement} combines WF and FF as follows: It begins like WF; once the load passes a threshold, the algorithm switches to an FF regime.
Throughout our evaluation, we used 0.6 as the threshold for the Adaptive algorithm.

The algorithm \emph{DistFromDiag}~\cite{RazPlacement} attempts to balance the host's resource consumption according to its proportions.
For example, if a host has 100GB disk and 10GB RAM, it aspires for a 10:1 ratio between available disk and RAM. 

We also consider two algorithms that incorporate randomization into WF and FF. These variants, referred to as \emph{WorstFit-Rand (WFR)} and \emph{FirstFit-Rand (FFR)}, respectively, weigh the hosts based on the WF and FF strategies but randomly select a host from the $\Lambda$ top-ranking available hosts (in the spirit of the option available in OpenStack, as described in Section~\ref{sec:background}).
In our evaluation of WFR and FFR, we set $\Lambda=5$. 

Finally, we evaluate the \emph{Random} algorithm, which selects a host uniformly at random among the available hosts.

\subsection{Datasets}
\label{Sec:Datasets}
We use three datasets that capture requests made in real systems. We evaluate each workload in a cloud environment with sufficiently many hosts to accommodate all the requests (see Section~\ref{sec:experiments} for details on choosing the number of hosts).

\textbf{NFV Dataset} was collected from a proprietary large NFV management and orchestration (MANO) system~\cite{ASC}. 
In this scenario, hosts are identical, and the placement requests are for VMs of preset sizes (flavors). Hosts and placement requests are two-dimensional tuples of the form $\left<memory, storage\right>$. 
The sizes are normalized such that hosts' capacity is $\left<1,1\right>$, and each VM requires a certain fraction of this capacity.
Table~\ref{table:proprietary} shows the distribution of flavors for this dataset. 

\begin{table}[tbh]
	\centering
    \caption{Normalized breakdown of requests for VM images by memory and storage, obtained from the NFV dataset.}
    \label{table:proprietary}
	\renewcommand\arraystretch{1.3}
    \TableFontSize
	\begin{tabular}{|cc||ccccc|c|}
		\hline
		& & \multicolumn{6}{c|}{$storage$} \\
		& & 0.01 & 0.04 & 0.1 & 0.3 & 0.54 & Total \\ 
		\hline
		\hline
		\parbox[t]{2mm}{\multirow{6}{*}{\rotatebox[origin=c]{90}{$memory$}}}
		& 0.001 & 14 & 22 & 14 & 3 & 13 & 66 \\
		& 0.016 &    7 &     93 &     0    & 2 &    0     & 102\\
		
		& 0.032 &    83 &    165 &    0 &    14    & 0 & 262 \\
		& 0.064 &    1    & 1 &     1 &    0    & 0 &    3 \\
		& 0.19    & 0&    2    &0 &    0&    2&    4\\
		\cline{2-8}
		& Total    & 105    & 283 &    15 &    19 &    15    & 437 \\
		
		\hline
	\end{tabular}

\end{table}

\textbf{Google Dataset}, recorded in a Google's cluster~\cite{clusterdata:Reiss2011}, holds data from 12,477 virtual machines characterized by tuples of $\left<CPU, memory\right>$. 
The normalized CPU values vary between $0.25$, $0.5$, and $1$, whereas the memory values can be grouped around
five levels:  $0.125$, $0.25$, $0.5$, $0.75$, and $1$~\cite{liu}. The hosts capacities are either  $\left<1,2\right>$ or $\left<2,1\right>$ in equal proportions~\cite{RazPlacement}.
Table~\ref{table:google} provides the breakdown of flavors for this dataset. 

\begin{table}[tbh]
	\centering
    \caption{Breakdown of the number of placement request sizes by CPU and memory, obtained from the Google dataset.}
    \label{table:google}
    \TableFontSize
    \renewcommand\arraystretch{1.3}
	\begin{tabular}{|cc||ccc|c|}
		\hline
		& & \multicolumn{4}{c|}{$CPU$} \\
		& & 0.25 & 0.5 & 1.0 & Total \\ 
		\hline
		\hline
		\parbox[t]{2mm}{\multirow{6}{*}{\rotatebox[origin=c]{90}{$memory$}}}
		& 0.125 & 0 & 60 & 0 & 60 \\
		& 0.25 & 123 & 3,835 & 0 & 3,958 \\
		& 0.5 & 0 & 6,672 & 3 & 6,675 \\
		& 0.75 & 0 & 992 & 0 & 992 \\
		& 1.0 & 0 & 4 & 788 & 792 \\
		\cline{2-6}
		& Total & 123 & 11,563 & 791 & 12,477 \\
		\hline
	\end{tabular}
\end{table}

\begin{table*}[t!]
	\centering
    \caption{Breakdown of placement request flavors of $\left<CPU,memory\right>$ obtained from the Amazon EC2 dataset. Flavors are sorted by $CPU$.}
    \label{table:amazon}
	\renewcommand\arraystretch{1.3}
    %\scriptsize
	\footnotesize
	
	\begin{tabular}{|c||c|c|c|c|c|c|c|c|c||c|c|c|c|c|c|c|c|c|}
	   % \hline
	    \cline{2-16}
	    \multicolumn{1}{c||}{} &
	    \multicolumn{9}{c||}{Small} &
	    \multicolumn{6}{c|}{Large} \\
		\hline
		\hline
		$CPU$ & 0.035 & 0.07 & 0.083 & 0.1 & 0.142 & 0.167 & 0.2 & 0.333 & 0.354 &    
		0.4 & 0.5 & 0.5 & 0.8 & 0.833 & 1
		\\ 
		$memory$ & 0.008 &    0.016 & 0.031 &  0.008 & 0.031  &    0.063      & 0.016 & 0.125 & 0.062 &    
		0.031   &  0.125 &    0.5 &   0.063  &    0.25 &     0.25  \\

		\hline
	\end{tabular}
\end{table*}

\textbf{Amazon Dataset} is based on data from Amazon EC2 hosts and VM flavors~\cite{Mills11comparingvmplacement,RazPlacement}. Table~\ref{table:amazon} depicts the
flavors of the normalized $\left<CPU, memory\right>$ in this dataset, 
where each column represents one possible flavor of requests. We partition requests' flavors into two types: {\em small} flavors, which have a CPU requirement below $0.4$, and {\em large} flavors, which consist of all remaining flavors.
We generate a sequence of $1000$ small requests and $100$ large
ones (i.e., a total of $1100$ requests) and select a flavor for each request
% u.a.r.
uniformly at random
from the corresponding flavor types.
% In evaluating the performance for this dataset,
In this scenario
we consider hosts with capacities of either $\left<1, 2\right>$, or $\left<2,1\right>$ in equal proportions (similarly to the host setup used in the Google dataset).

\begin{figure}[ht]
\centering
\includegraphics[width=0.7\columnwidth] {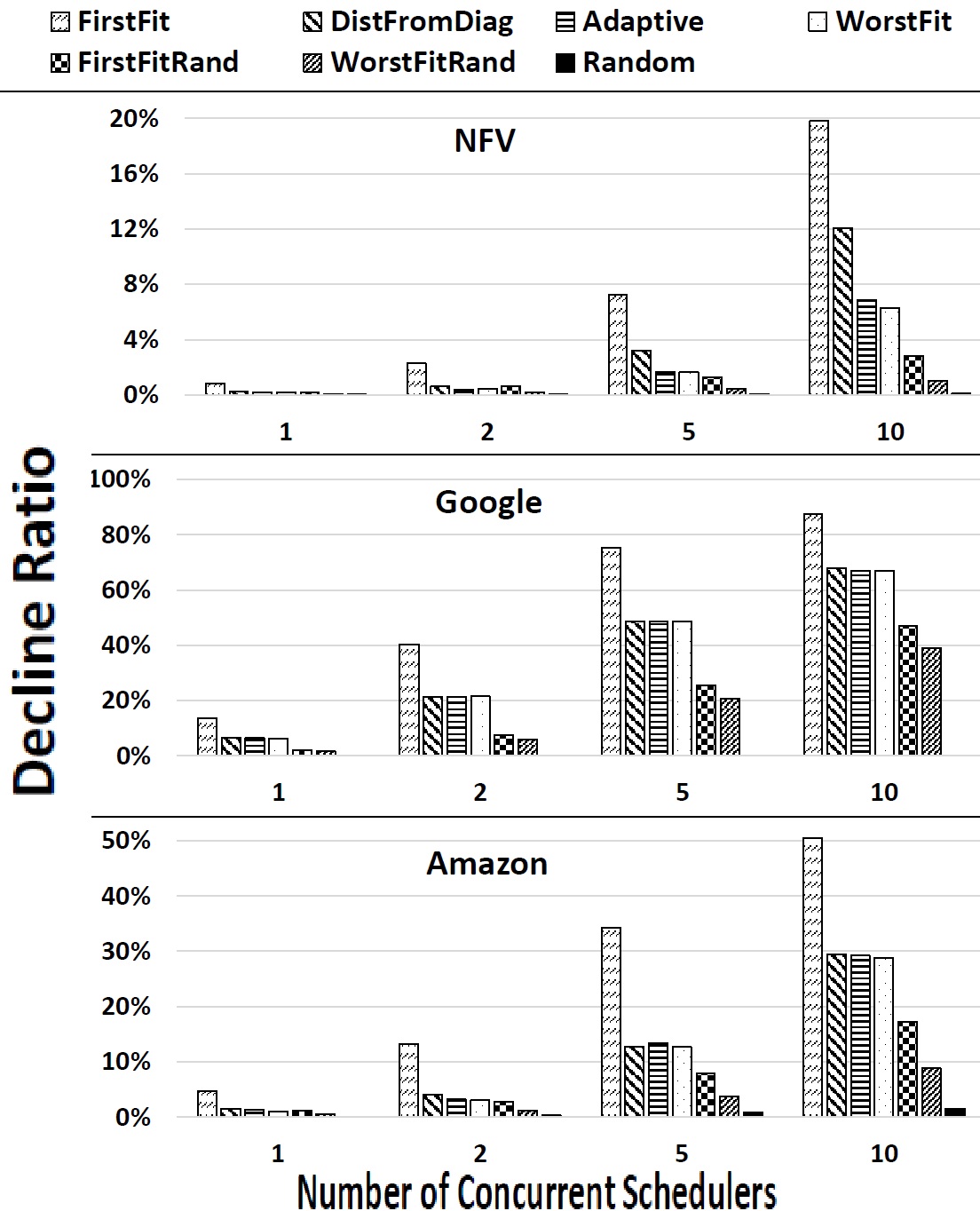}
\caption{Decline ratios for different placement algorithms and a varying number of parallel schedulers on the NFV, Google, and Amazon datasets. Note that the decline ratio (y-axis) ranges corresponding to the various datasets are distinct.} 
\label{fig:sim_random_rules}
\vspace{-0.25cm}
\end{figure}

\subsection{Experiments}
\label{sec:experiments}
We now turn to study the effect of running multiple parallel schedulers with existing algorithms.
We select a number of hosts that enable placing all requests at once (by some algorithm). 
Evaluating the required number of hosts to accommodate all the requests in a given trace is equivalent to the multi-dimensional bin packing problem, which is NP-hard~\cite{garey79computers}. Thus, we approximate this number as suggested in~\cite{RazPlacement}: We run the trace for each algorithm multiple times, each time with a randomly generated order of requests.
Whenever the placement algorithm fails to accommodate a request with the currently available resources, we open a new host.
The approximated value is the minimal number of open hosts in all runs. %for the required number of hosts is the minimum obtained overall algorithms and orders of requests.

To simulate large clouds, we replicated the NFV dataset to have 4730 requests with 279 hosts. The Amazon dataset is evaluated with 126 hosts, and the Google dataset with 5989 hosts.
% Throughout 
Our experiments make just one attempt to place any request (i.e., we do not retry placing declined requests). 

Our results are illustrated in Fig.~\ref{fig:sim_random_rules}.
When using a single scheduler, there are very few failures in all policies.
% \gabi{this is not very precise. FF seems to do pretty bad even for one scheduler, most notably in the Google dataset. Any idea why is that?}
% \gabi{check first fit vs. best fit throughout}
Yet, the decline ratio in Random remains low also for higher levels of parallelism.
This result is intuitive as randomly allocating requests to hosts minimizes the probability of having many schedulers select the same host concurrently.
In contrast, FirstFit is the worst, as all the schedulers select the same host even if it is close to being complete.
In other algorithms like WorstFit, once a host is nearly full, it is less attractive, and thus the schedulers distribute their placement decisions upon a larger number of hosts. 

The decline ratio of the deterministic algorithms becomes very high, even when running only $10$ schedulers. 
This problem is somewhat mitigated by OpenStack's solution of introducing slight randomization into traditional algorithms (as captured by FFR and WFR). However, statically setting $\Lambda = 5$ is insufficient when having ten schedulers. These results show that the OpenStack community correctly identified the problems with parallelism and introduced a valid workaround. However, the interplay between parallelism and decline ratio has not been studied. 
Our work builds upon the insights drawn from the above results and claims that one should use randomness to maximize parallelism in resource management.
In particular, our goal is to study the scaling laws of parallelism when combined with random VM placement. 

\section{Adaptive Partial State Random (APSR)}
\label{sec:APSR}
This section presents our algorithm \emph{Adaptive Partial State Random (\algtop)}. Motivated by our observations from Section~\ref{sec:PlacementStudy}, \algtop\ implements an efficient random policy that dynamically adjusts the number of schedulers ($s$) according to the system's perceived utilization. Whenever \algtop\ uses parallel schedulers ($s>1$), it is guaranteed to satisfy the SLA and budget constraints.

Upon receiving a placement request, each \algtop\ scheduler does the following:
\begin{inparaenum}[(i)]
\item queries $d$ hosts (for some value $d$), 
\item filters out hosts that cannot accommodate the request,
\item randomly selects an available host out of the remaining set of hosts, and
\item sends the request to the chosen host.
\end{inparaenum}

\algtop\ relies on a centralized controller called the \emph{\algtop\ controller} to do the following periodically:
\begin{inparaenum}[(i)]
\item estimate the system's utilization, captured by the estimate $\nfreebins$ of the number of available hosts,
\item determine the number $\nsched$ of parallel schedulers, and
\item determine the number $d$ of hosts each scheduler queries per request. 
\end{inparaenum}
The controller determines the above parameters to ensure the validity of the configuration.

\begin{algorithm}[t!]
\footnotesize
\caption {\algtop\ Controller ($\nbins, \dr, \budget, T$)} \label{alg:algtop}
\begin{algorithmic}[1]
\State $\nsched \gets 1$, $k \gets n$ \label{alg:algtop:nsched_gets_1}
\State GenerateSchedulers($1, \budget$)\label{alg:algtop:Generate_1_sched}
\For{every time slot $t=T,2T,3T,\ldots$}\label{alg:algtop:While_loop_start}
    \State \label{alg:APSR:estimateK} $\nfreebins \gets$ \estkalg($\ldots$)
    \State ($\nsched, \dqueries$) $\gets$ \maxparalg($\nbins, \dr, \budget, \nfreebins$)
    \label{alg:algtop:call_max_paral}
        \State GenerateSchedulers($\nsched, \dqueries$)
\EndFor
\label{alg:algtop:While_loop_end}
\end{algorithmic}
\end{algorithm}

Algorithm~\ref{alg:algtop} illustrates the \algtop\ controller algorithm. 
The procedure GenerateSchedulers($\nsched, d$) adjusts the number of schedulers to $\nsched$ and the number of hosts queried by each scheduler to $d$.
The method \estkalg\ estimates the number of hosts $\nfreebins$ that can accommodate a request. We do not specify the arguments for this method since it can be implemented in various ways (see details in~Section~\ref{sec:practical_implementation}).
The procedure \maxparalg\ considers the system state and the SLA constraints and outputs the number of schedulers $\nsched$ and the number of hosts each scheduler queries ($d$). 

\section{Analysis}\label{sec:comb_analysis}
We now establish the correctness of our approach. 
We start with a simplified {\em balls-and-bins} model where hosts are unit-size {\em bins}, and requests are unit-size {\em balls}, implying that each bin can store at most one ball.
Each scheduler is an {\em agent} assigning balls to bins.
We show sufficient conditions for satisfying the SLA requirement in this simplified model. Our conditions provide a lower bound on the number of parallel agents for a given failure probability. We further show that the decline ratio serves as an upper bound on the original model's decline ratio. These results imply that when \algtop\ utilizes parallelism, the decline ratio is at most $\dr$, and the total number of queries performed by all agents is at most $B$. 

\subsection{Balls-and-bins Model}
\label{SubSec:balls_n_bins_model}
Assume $\nsched$ identical agents acting in parallel,  trying to place balls in available bins.
Each agent queries $d$ random bins and possibly finds some of them available.
If the agent does not find any available bins, the ball assignment fails. Otherwise, the agent selects an available bin uniformly at random and tries to place its ball in that bin.

Agents are unaware of the decisions made by other agents, which may cause multiple agents to select the same available bin. In such a case, one of the agents succeeds, and the rest of them fail.
We use the term {\em potentially happy agent} to refer to an agent that finds an available bin.
Similarly, the term {\em happy agent} refers to an agent that successfully places a ball in an available bin. Finally, we use the term {\em unhappy agent} to refer to an agent that fails to place its ball (either due to collision or due to not finding an available bin).

We let the random variables, $\npothappy$ and $\nhappy$, denote the number of potentially happy agents and happy agents.
We denote by $\nfreebins$ a lower bound on the number of available bins in some time slot where agents contend for assigning balls into bins.

We view the SLA requirement of having a decline ratio of at most $\dr$ as a lower bound on the probability that an arbitrary agent attempting to assign a ball to some bin is happy. Formally, this requirement translates to ensuring that:  
\insertShortSection {
    $\frac{E[\nhappy]}{\nsched} \geq 1 - \dr.$
}
\insertFullSection {
    \begin{align} \label{Eq:SLA_1}
    \frac{E[\nhappy]}{\nsched} \geq 1 - \dr.
    \end{align}
}

We also require that the total number of bins queried by our agents is no more than a prescribed budget ($B$), which translates to requiring that:
$
\nsched \cdot d \leq \budget.
$

Given $\nbins, \nfreebins, \dr$, and $\budget$, our goal is to find the largest number of agents $\nsched$, and the number of bin queries per agent $d$, that satisfy the above conditions. 

We calculate the expected number of happy agents $E[\nhappy]$ in order to estimate the failure probability.  
Observe that  $E[\nhappy]$ can be expressed by conditioning the number of happy agents $\nhappy$ on the number of potentially happy agents $\npothappy$. I.e., 
\begin{equation}\label{Eq:Ehs_by_Ehs_cond_f}
   E[\nhappy] = 
   \sum_{f=1}^{s} \bigg[\Pr (\npothappy = f) \cdot E[\nhappy | \npothappy = f] \bigg].
\end{equation}

We now turn to evaluate the probability distribution of 
$\npothappy$, and then calculate the conditional expectation $E[\nhappy | \npothappy = f]$. 

To evaluate the distribution of the number of potentially happy agents  $\npothappy$, observe that an agent fails to find an available bin with probability
$\left (\frac{\nbins-\nfreebins}{\nbins} \right)^d$. Therefore, the probability that an agent is potentially happy is: 
\begin{equation}\label{Eq:def_sigma}
\PrFF = 1 - \left (\frac{\nbins-\nfreebins}{\nbins} \right)^d.     
\end{equation}

One can interpret 
$\npothappy$ as the result of $\nsched$ independent Bernoulli trials with success probability $\PrFF$. Therefore:
\begin{equation}\label{Eq:Fs}
Pr (\npothappy = f) = Bin (f, \nsched, \PrFF),
\end{equation}

where 
\begin{equation}\label{Eq:def_bin}
Bin (f, \nsched, \PrFF) \equiv 
\binom {\nsched}{f}  \PrFF^f (1-\PrFF)^{\nsched-f}.
\end{equation}

For calculating $E[\nhappy | \npothappy = f]$, we examine the process of the potentially happy agents placing their balls from the point of view of the $\nfreebins$ free bins. For ease of presentation, we associate 
each potentially happy agent with a sequence number $1, \dots, f$, 
and each available bin with a sequence number $1, \dots, \nfreebins$. 

The following proposition
shows that the probability that an arbitrary potentially happy agent selects an arbitrary available bin is uniform over all available bins.

\begin{proposition}\label{Prop:one_over_k}
If agent $i$ is potentially happy, then it places its ball on available bin $j$ with a probability of $\frac{1}{k}$.
\end{proposition}

\insertproofsketch{
\begin{proof}[Proof sketch]
The proof follows from
considering
the conditional probability of an agent finding a specific available bin $j$, conditioned on having found $x$ available bins in its set of sampled bins. This probability is $\frac{x}{k}$ due to the uniformity of the agent's choice. Selecting any specific available bin $j$ has the same probability, which implies that the probability for finding a specific bin is $\frac{1}{k}$, as required.
\end{proof}
}

% ===========================
% INSERT PROOF - START
% ===========================
\insertproof{
\begin{proof}
Let $i$ be an agent -- not necessarily a potentially happy agent.
Denote by $Q_i$ the set of bins which agent $i$ queries and finds available. 
Let $q_i$ denote the random variable for the number of bins which agent $i$ finds available, namely, $\abs{Q_i} = q_i$. Then 
$\Pr (q_i = x)$ captures the probability that agent $i$ finds $x$ distinct available bins in his overall $d$ samples. 

For each available bin $\ell$, we let  $B_\ell$ denote a binary random variable, indicating whether agent $i$ samples bin $\ell$. Namely, $B_\ell=1$ iff $\ell \in Q_i$. Then we have
\begin{equation}\label{Eq:indicator_rv}
\begin{split}
\sum_{\ell=1}^k \Pr \left(\ell \in Q_i | q_i = x \right) & = 
\sum_{\ell=1}^k \Pr(B_\ell = 1 | q_i = x) \\
& = 
\sum_{\ell=1}^k E\left[B_\ell | q_i = x\right] \\
& = 
E \left[ \sum_{\ell=1}^k B_\ell | q_i = x \right] \\
& = 
E [q_i | q_i=x] = x.
\end{split}
\end{equation}

Since agent $i$ samples the bins i.i.d., we have for each  available bins $\ell, \ell'$ that $\Pr (\ell \in Q_i | q_i = x) = \Pr (\ell' \in Q_i | q_i = x)$. 
By~\eqref{Eq:indicator_rv} it follows that 
for every $\ell = 1, \dots, \nfreebins$, 
$\Pr (\ell \in Q_i | q_i = x) = \frac{x}{\nfreebins}$.

Hence, the probability that agent $i$ selects available bin $j$ is
\begin{equation}\label{Eq:i_sel_j}
\begin{split}
\Pr (j \in Q_i) 
& = 
\sum_{x=1}^{\nfreebins} \Pr (q_i = x) \cdot
\Pr (j \in Q_i | q_i = x) \\
& =
\frac{1}{\nfreebins}
\sum_{x=1}^{\nfreebins} x \cdot \Pr (q_i = x).
\end{split}
\end{equation}

If agent $i$ samples available bin $j$, then she selects $j$ w.p. $\frac{1}{x}$. It follows that
\begin{align}\label{Eq:i_smpls_j}
\begin{split}
\Pr (i \ \textrm{selects} \ j) 
& =
\frac{1}{\nfreebins}
\sum_{x=1}^{\nfreebins} x \cdot \Pr (q_i = x) \cdot 
\frac{1}{x} \\
& = 
\frac{1}{\nfreebins}
\sum_{x=1}^{\nfreebins} 
\Pr (q_i = x). 
\end{split}
\end{align}

Observe that agent $i$ is potentially happy iff she samples at least one available bin, that is, if $q_i > 0$. The probability for this event is 
$\sum_{x=1}^{\nfreebins} \Pr (q_i = x)$. Combining this observation with~\eqref{Eq:i_smpls_j}, the probability that agent $i$ samples available bin $j$ given that $i$ is potentially happy is $\frac{1}{k}$.
\end{proof}
}
% ===========================
% INSERT PROOF - END
% ===========================

By Proposition~\ref{Prop:one_over_k}, the probability that potentially-happy agent $i$ does \emph{not} place its ball in bin $j$ is $1 - \frac{1}{\nfreebins} = \frac{k-1}{k}$. As the agents are mutually independent, the probability that none of the $f$ potentially happy agents places its ball in bin $j$ is $\left( \frac{k-1}{k} \right)^f$. The probability that at least one of the $f$ potentially happy agents tries to place its ball in bin $j$ is $1 - \left( \frac{k-1}{k} \right)^f$. 
From the point of view of bin $j$, this process is equivalent to a Bernoulli trial, which succeeds iff at least one agent places its ball in bin $j$.
If 
% the Bernoulli 
this
% trial 
succeeds, bin $j$ is exclusively associated with a single happy agent.

Applying the analysis above for each of the $\nfreebins$ free bins, we obtain that $E[\nhappy | \npothappy = f]$ is equivalent to the expected number of successes in $\nfreebins$ independent Bernoulli trials, with probability of success $1 - \left( \frac{k-1}{k} \right)^f$ each. Hence, 

\begin{equation}\label{Eq:EHs_cond_f_correct}
E[\nhappy | \npothappy = f] = \nfreebins \left[
1 - \left( \frac{k-1}{k} \right)^f
\right].
\end{equation}
Combining~\eqref{Eq:Ehs_by_Ehs_cond_f} with~\eqref{Eq:Fs} and~\eqref{Eq:EHs_cond_f_correct}, we obtain
% the following  expression for $E[\nhappy]$
\begin{equation}\label{Eq:EHs}
E[\nhappy]  = 
\nfreebins 
\sum_{f=1}^{\nsched} 
\left[
1 - \left( \frac{k-1}{k} \right)^f
\right]
\cdot
Bin (f, \nsched, \PrFF).
\end{equation}

The following corollary is a direct consequence of~\eqref{Eq:SLA_1} and~\eqref{Eq:EHs}.
% provides a sufficient condition for fulfilling the SLA. 
\begin{corollary}\label{cor:balls_n_bins_SLA}
If 
$
\nfreebins 
\sum_{f=1}^{\nsched} 
\left[
1 - \left( \frac{k-1}{k} \right)^f
\right]
\cdot
Bin (f, \nsched, \PrFF)
\geq \nsched (1 - \dr)$
then the expected decline ratio with
$\nsched$ agents, where each agent queries $\dqueries$ bins, is at most $\dr$.
\end{corollary}

Based on Corollary~\ref{cor:balls_n_bins_SLA}, we now describe the details of the \maxparalg\ method, which maximizes the parallelism while satisfying the SLA and budget constraints. The method is detailed in Algorithm~\ref{alg:max_paral}.
After initially setting
% the number of schedulers to 
$\nsched=1$, the algorithm repeatedly increases the value of $\nsched$, while maintaining feasibility by having
SatisfySLA validate that the condition of Corollary~\ref{cor:balls_n_bins_SLA} is satisfied for the given configuration.

\begin{algorithm}[t!]
\footnotesize
\caption {\maxparalg \ ($\nbins, \dr, \budget, \nfreebins$)} \label{alg:max_paral}
\begin{algorithmic}[1]
\State $\nsched \gets 1$\label{alg:max_paral:line_s_eq_1}\Comment{initialization} 
\While {SatisfySLA $\left(\nbins, \dr, \nfreebins, \nsched+1, \left \lfloor{\frac{\budget}{\nsched+1}} \right \rfloor \right)$}\label{alg:max_paral:while_s_begin}
    \State $\nsched \gets \nsched + 1$
\EndWhile \label{alg:max_paral:while_s_end}
\State return $\nsched, \big\lfloor{\frac{\budget}{\nsched}} \big\rfloor$ \label{alg:max_paral:while_d_end}
\end{algorithmic}
\end{algorithm}

\subsection{SLA Guarantees with Availability Lower Bounds}
% We now show that the value $\nfreebins$ is a {\em lower bound} on the number of available hosts for any request.
% which gives us  an upper bound on the system's utilization.
We 
% begin by showing 
first show
that if $\nfreebins$ is the precise number of available hosts for any request, then \maxparalg\ indeed generates a valid configuration.

\begin{theorem}\label{thm:practic_alg_SLA}
Assume $k$ is the number of available hosts that may accommodate any request flavor.
If \maxparalg($\nbins, \dr, \budget, \nfreebins) = (\nsched, d)$ and $\nsched>1$ then employing $\nsched$ schedulers, each querying $\dqueries$ hosts,
% during time interval $T$
guarantees an expected decline ratio of at most $\dr$.
% \itamar{The condition $\nsched>1$ is because  $\nsched=1$ is the default value. We don't use $\nsched=0$ as default, because this will force the upper-level alg' to check the value returned by \maxparalg, to ensure at least 1 scheduler is generated.}
\end{theorem}
% \itamar{I changed here to proof sketch to reduce space. Please revise it.}
\insertproofsketch{
\begin{proof}[Proof sketch]
Our proof will consider the following {\em compacting} process:
\begin{inparaenum}[(i)]
\item \label{condense:1} Consider all the hosts in $\hosts_{\flavor^*}$ as available for all flavors,
\item \label{condense:2} consider the other hosts as unavailable for any request, and 
\item \label{condense:3}determine that once a scheduler allocates a request in a host, it becomes unavailable. 
\end{inparaenum}   
The compacted system is equivalent to our balls-and-bins model. 
Hence, by Corollary~\ref{cor:balls_n_bins_SLA}, the compacted system satisfies the SLA constraint. 
One can then show that the decline ratio in the compacted system serves as an upper bound on the decline ratio ($\dr$) in the original system. The result follows.
\end{proof}
}

\insertproof{
\begin{proof}
Let $\hosts_{\flavor}$ denote the set of hosts with enough resources for accommodating a request of flavor $\flavor$. Using our notation, it follows that $\abs{\hosts_{\flavor}} = \nfreebins_{\flavor}$.
Let $\flavor^* = \arg\min_{\flavor}\set{\nfreebins_{\flavor}}$.

Consider the following {\em compacting} process:
\begin{enumerate}
\itemsep-0.2em 
    \item \label{condense:1} Consider all the hosts in $\hosts_{\flavor^*}$ as available for all flavors.
    \item \label{condense:2} Consider the other hosts as unavailable for any request.
    \item \label{condense:3} Determine that once a scheduler allocates a request in a host, it becomes unavailable. 
\end{enumerate}   

We claim that compacting the system can only increase its decline ratio for the following reasons:
First, as for each $\flavor \in \flavors$ we have $\nfreebins_{\flavor^*} \leq \nfreebins_{\flavor}$, steps~\ref{condense:1} and~\ref{condense:2} can only decrease the number of hosts available for each flavor. This reduces the expected number of available hosts found by each scheduler.
Second, steps~\ref{condense:1} and~\ref{condense:2} define the available hosts of any flavor to be exactly $\hosts_{\flavor^*}$. This compacting may only increase the probability that multiple schedulers will end up assigning their requests to the same host.
Finally, a host may accommodate multiple parallel requests providing it has enough resources while step~\ref{condense:3} disallows it, which implies a potential increase in the decline ratio.
Thus, any algorithm satisfying the SLA in the compacted system also satisfies it in the original system.

We now note that the compacted system is equivalent to our balls-and-bins model. To see this, observe that once the sets of available hosts for every request become identical (due to steps~\ref{condense:1} and~\ref{condense:2}), the requests themselves are also virtually identical and thus become equivalent to the identical balls in our balls-and-bins model. Furthermore, as every host can accommodate only a single request (due to step~\ref{condense:3}), the hosts can be modeled as identical bins, where each available bin can accommodate merely a single ball.

By Corollary~\ref{cor:balls_n_bins_SLA}, 
\maxparalg\ satisfies the SLA requirement in the balls-and-bins model, which is equivalent to guaranteeing SLA also in the compacted system. As the decline ratio in the compacted system serves as an upper bound on the decline ratio ($\dr$), the result follows.
\end{proof}
} % end \insertproof

The proof of Theorem~\ref{thm:practic_alg_SLA} implicitly suggests that all the requests are handled in a time slot belonging to the flavor with the minimum number of available hosts. Furthermore, it suggests that two requests can never be placed in parallel on the same host. Thus, we expect better decline ratios in practice. 

The following corollary
shows that for providing performance guarantees, it is sufficient to know only a {\em lower bound} on the number of hosts available for every request flavor.

\begin{corollary}
\label{cor:thm_holds_whenever_k_is_lower_bound}
Theorem~\ref{thm:practic_alg_SLA} holds whenever $k$ is a lower bound on the number of available hosts for every request flavor.
\end{corollary}

\insertproofsketch{
\begin{proof}[Proof sketch]
The proof is based on showing that increasing the number of hosts available for every request flavor while keeping the number schedulers $\nsched$ and the sample size $d$ unchanged does not increase the decline ratio.
We introduce a $k$ superscript to our various notations to indicate the values for a specific value of $k$.
We then rearrange~\eqref{Eq:Ehs_by_Ehs_cond_f}, capturing the expected number of happy agents, and show that it can be cast as
% \begin{equation}
$
E\left[\hk\right]
= \sum_{f=0}^{s-1} \left[\Pr (\phk > f) \cdot \dif (k, f) \right]
$,
% \end{equation}
where we let
% \begin{equation}
$
\dif (k, f) = E \left[\hk | \phk = f+1\right] - 
E \left[\hk | \phk = f\right]
$.
% \end{equation}
It can be shown that both $\Pr (\phk > f)$ and $\dif (k, f)$ are monotone non decreasing in $k$, thus completing the proof.
\end{proof}
}

% ===========================
% INSERT PROOF - START
% ===========================
\insertproof{
\begin{proof}
We have to show that increasing 
the number of hosts available for every request flavor, 
while keeping the number schedulers $\nsched$ and the sample size $d$ unchanged can only decrease the decline ratio. 
We do so by checking the effect of increasing 
the number of available bins $\nfreebins$ 
on our balls-and-bins analysis.
As we now vary 
$\nfreebins$,
we add to the notation of our random variables a superscript indicating its value.
That is, $\phk$ and $\hk$ denote the random variable for the number of potentially happy and happy agents, respectively,  when there are $\nfreebins$ available bins. 
Recalling the SLA requirement in~\eqref{Eq:SLA_1}, it suffices to show that $E\left[\hkpp \right] \geq E\left[ \hk \right]$.

Using our modified notation, we rewrite~\eqref{Eq:Ehs_by_Ehs_cond_f} as
\begin{equation}\label{Eq:Ehs_by_Ehs_cond_f_k}
  E\left[\hk\right] = 
    \sum_{f=1}^{s} \Pr \left(\phk = f \right) \cdot E \left[\hk | \phk = f \right] 
\end{equation}

We now handle  each of the components in the product appearing on the right-hand side of~\eqref{Eq:Ehs_by_Ehs_cond_f_k} separately, namely
\begin{inparaenum}[(i)]
\item the probability distribution of the number potentially happy agents, and
\item the expected number of happy agents, given that there are $f$ potentially happy agents.
\end{inparaenum} 

Intuitively, the probability of having more than $f$ potentially happy agents is non-decreasing in the number of free bins $\nfreebins$. Indeed, 
combining~\eqref{Eq:def_sigma},~\eqref{Eq:Fs} and~\eqref{Eq:def_bin} shows that 
\begin{equation}\label{Eq:Fs_inc_in_k}
\Pr \left(\phkpp > f\right) \geq \Pr \left(\phk > f\right).
\end{equation}

To quantify the impact of the number of potentially happy agents $f$ on the expected number of happy agents  we
let $\dif(k, f)$ denote the difference function
\begin{equation}\label{Eq:def_d}
\dif (k, f) = E \left[\hk | \phk = f+1\right] - 
E \left[\hk | \phk = f\right].
\end{equation}

$\dif(k, f)$ captures the contribution of adding one potentially happy agent to the expected number of happy agents. 
As $E [\hk | \phk = 0] = 0$, we have $\dif (k, 0) = E [\hk | \phk = 1]$. We can therefore rewrite~\eqref{Eq:Ehs_by_Ehs_cond_f_k} as follows:
\begin{equation}\label{Eq:EHs_by_f_and_dif}
\begin{split}
E \left[\hk \right] 
= & \sum_{f=1}^s \Pr(\phk = f) \cdot E \left[\hk | \phk = f \right] \\
= & \Pr(\phk = 1) \cdot \dif (k, 0) \ + \\
& \Pr(\phk = 2) \cdot \left[\dif (k, 0) + \dif (k, 1) \right] + \dots + \\
& \Pr(\phk = s) \cdot \\
& \left[\dif (k, 0) + \dif (k, 1) + \dots + \dif (k, s-1) \right]\\
= & \sum_{f=0}^{s-1} \left[\Pr (\phk > f) \cdot \dif (k, f) \right]
\end{split}    
\end{equation}

By combining~\eqref{Eq:Fs_inc_in_k} and~\eqref{Eq:EHs_by_f_and_dif}, it suffices to show that
\begin{equation}\label{Eq:d_inc_in_k}
\dif(k+1, f) > \dif (k, f)    
\end{equation}

For proving  that~\eqref{Eq:d_inc_in_k} is satisfied, we assign~\eqref{Eq:EHs_cond_f_correct} in the definition of $\dif(*)$ in~\eqref{Eq:def_d}, and obtain:
\begin{equation}\label{Eq:dif_k_correct}
\begin{split}
\dif (k, f) & = 
\nfreebins \left[
\left( \frac{k-1}{k} \right)^{f}
- 
\left( \frac{k-1}{k} \right)^{f+1} 
\right]  \\
& = 
\left( \frac{k-1}{k} \right)^f
\end{split}
\end{equation}

Hence,
\begin{equation}
\begin{split}
\dif (k+1, f) - \dif (k, f) = 
\left( \frac{k}{k+1} \right)^f -
\left( \frac{k-1}{k} \right)^f
\geq 0
\end{split}
\end{equation}
where the last inequality is satisfied for every $\nfreebins > 0$.
\end{proof}
}
% ===========================
% INSERT PROOF - END
% ===========================

\section{Practical Implementation of \algtop}
\label{sec:practical_implementation}
We now discuss practical aspects of implementing \algtop.
The main caveat in implementing \algtop\ is to estimate the number of available hosts for any request flavor ($k$).

A straightforward option is to compute $\nfreebins$ explicitly by running a centralized periodic task that gathers the state from all hosts. We note that the \algtop\ controller may execute such a task (in Line~\ref{alg:APSR:estimateK}). When the task is performed every time step (i.e., by setting $T=1$ in \algtop), then the guarantees of Theorem~\ref{thm:practic_alg_SLA} hold.
However, this approach incurs the communication overhead of querying all the hosts.%\footnote{Note that all the alternative placement algorithms considered in section~\ref{sec:PlacementStudy} have {\em each scheduler} query all the hosts in every time slot.}

Alternatively, we propose estimating $\nfreebins$ by relying on the statistics the schedulers gather during their regular operation. Algorithm~\ref{alg:estimateK} describes our proposed algorithm \estkalg($k$) for estimating $\nfreebins$.

\begin{algorithm}[t!]
\footnotesize
\caption {\estkalg($k$)} \label{alg:estimateK}
\begin{algorithmic}[1]
    \ForAll{$\flavor \in \flavors$} \Comment{for each flavor}
        \State $\nbins_{\flavor}^{(tot)} \gets \sum_{i=1}^{\nsched} \nbins_{\flavor}^{(i)}$ ,$\nfreebins_{\flavor}^{(tot)} \gets \sum_{i=1}^{\nsched} \nfreebins_{\flavor}^{(i)}$
        \label{alg:algtop_n_i_tot}
        %\Comment{total number of queries for requests of flavor $i$}
        \label{alg:algtop_k_i_tot}
    \EndFor
    \State $\tilde{\nfreebins} \gets
    \nbins \cdot \min_{\flavor \in \flavors}
    \left[ \frac{\nfreebins_{\flavor}^{(tot)}}{\nbins_{\flavor}^{(tot)}} \right]$ \label{alg:algtop:assign_k_n}
    \State return $\alpha \cdot  \tilde{\nfreebins} + (1 - \alpha) \cdot \nfreebins$ \label{alg:algtop:exp_moving_average}
\end{algorithmic}
\normalsize
\end{algorithm}

Our algorithm assumes that each scheduler $i$ maintains counters $\nbins_{\flavor}^{(i)}$ and $\nfreebins_{\flavor}^{(i)}$, which keep track of the overall number of hosts queried, and the total number of available hosts of flavor $\flavor$, respectively.
These counters are reset before each call to algorithm \estkalg.
The algorithm uses these counters to estimate the {\em overall} number of hosts queried and the overall number of available hosts for each flavor.
These values can be used to estimate the {\em percentage} of hosts available for each request flavor.
The normalized minimum of all flavors is chosen as the pessimistic estimate of $k$. We then use exponential averaging to produce an updated estimate of $k$.
%The intuition underlying this estimation is that schedulers operate {\em independently}, and each scheduler queries a relatively small number of hosts in each time slot.
% The estimation is then combined with the concepts used for proving Theorem~\ref{thm:practic_alg_SLA}.

We emphasize that our approach does not require any additional querying of hosts.
We note that Algorithm~\ref{alg:estimateK} does not ensure that our estimate is a lower bound on the available resources in the system, as is required by Corollary~\ref{cor:thm_holds_whenever_k_is_lower_bound}. However, due to the conservative approach in making the estimate (namely, Line~\ref{alg:algtop:assign_k_n} in Algorithm~\ref{alg:estimateK}) our estimation method is effective when incorporated within our \algtop\ Algorithm.

\section{\algtop\ Evaluation}
\label{sec:sim_algtop}
This section positions \algtop\ with respect to known placement algorithms and evaluates
the interplay between parallelism, utilization, decline ratio, and throughput.

\insertShortSection{\textbf{Trace-based Simulation: }} % In the short version, this is only a paragraph
\insertFullSection{\subsection{Simulation Settings}} % In the journal version, this is a subsection
% \itamar{Reviewer comment for Infocom'21: What simulator was used (CloudSim? ICanCloud? Something the authors created themselves?), and what are its assumptions? Presumably the authors assume homogeneous hardware throughout the cloud, and no application-specific SLAs?}
We model the arrival of requests using a Poisson process with parameter $\lambda_a$. Unless stated otherwise, we set $\lambda_a$ to 20, and $\dr$ (\algtop's target decline ratio) to 5\%. 
We set \algtop's query budget to be $\budget = \nbins$. That is, the {\em overall} number of samples made by all of our parallel schedulers is the same as the number of samples done by {\em a single} OpenStack scheduler. 
We set \algtop's time interval for estimating the state of the cloud to be $T=10$ and set $\alpha=0.1$ for the \estkalg\ method. 

%Recall that the number of queries per request by any of the standard algorithms is $\nbins$. In contrast, \algtop\ complies a budget constraint $\budget$. Unless stated otherwise, we set $\budget = \nbins$. %That is, the total number of queries per time slot, by all schedulers together, is at most $\nbins$.
% , and $\budget$ (\algtop's communication budget) to the number of hosts ($\nbins$). 

We consider requests of unbounded duration as it is a common (though somewhat unrealistic) benchmark for placement algorithms~\cite{ASC}.
These settings provide a clear demonstration of the relationship between utilization and parallelism.
Due to space constraints, we omit our simulation results for finite duration requests but note that these results have similar qualitative characteristics for such settings.

We use the workloads described in Section~\ref{Sec:Datasets}, and simulate large clouds with 30 replicas of the NFV dataset, seven replicas of the Amazon dataset, and one replica of the Google dataset, attaining a total of 13110, 7700, and 12477 requests, respectively.
We determine the number of hosts as the number of hosts needed for successfully placing all the requests at once (by some offline algorithm), as described in Section~\ref{sec:experiments}; we use 837 hosts for NFV, 876 hosts for Amazon, and 5989 hosts for Google. 
As discussed in Section~\ref{sec:experiments}, for every algorithm, we make a single attempt to place each request and compute the decline ratio accordingly.

\subsection{Comparing \algtop\ to other algorithms}
\begin{table*}[t!]
	\centering
    \caption{Decline ratios (in \%, lower is better) of \algtop\ and other placement algorithms when varying the (fixed) number of schedulers ($\nsched$, higher is better). \algtop's throughput, captured by the average number of active schedulers ($\bar{\nsched}$), is listed below its decline ratio.}
    \label{table:aspr_quality}
	\renewcommand\arraystretch{1.3}
    \TableFontSize
	\begin{tabular}{|c|c|c|c|c|c|c|c|c|c|c|c|}
		\hline
	    Dataset & \textit{s} &  APSR & \rand & FF & FFR & WF & WFR & Diag & Adapt\\ 
		\hline
		\hline
		\parbox[t]{2mm}{\multirow{5}{*} {\rotatebox[origin=c]{90}{NFV}}}
		& 1 &   & 0.3 & 0.0 & 0.0 & 0.3 & 0.3 & 0.7 & 0.3  \\
		& 5 & 0.4 & 0.4 & 11.1 & 2.5 & 4.0 & 1.0 & 5.3 & 2.2  \\
		& 10 &$\bar{\nsched}$ = 14&  0.5 & 23.3 & 5.2 & 8.2 & 2.1 & 7.8 & 3.1  \\
		 & 20 & & 0.7 & 35.7 & 10.0 & 12.1 & 3.3 & 11.7 & 11.6 \\
		 
		& 50 &  & 0.8 & 39.0 & 10.8 & 16.7 & 3.9 & 16.4 & 16.0
		\\
		\hline
		\parbox[t]{2mm}{\multirow{5}{*} {\rotatebox[origin=c]{90}{Google}}}
		& 1  &    & 2.3 & 0.4 & 1.3 & 8.7 & 8.7 & 2.2 & 8.7 \\
		& 5  & 3.1  & 2.4 & 56.2 & 15.5 & 42.0 & 16.4 & 42.7 & 42.0 \\
		& 10 & $\bar{\nsched}$ = 20   & 2.4 & 77.8 & 29.9 & 64.1 & 26.1 & 62.7 & 64.5 \\
		& 20 & &   2.4 & 87.8 & 48.1 & 79.8 & 36.4 & 73.8 & 79.3  \\
		& 50 &  & 2.4 & 88.9 & 51.4 & 81.2 & 40.2 & 76.7 & 81.2 \\

\hline
		\parbox[t]{2mm}{\multirow{5}{*} {\rotatebox[origin=c]{90}{Amazon}}}
		& 1  &  & 0.5 & 0.0 & 0.0 & 0.4 & 0.4 & 1.3 & 0.2  \\
		& 5  & 0.8 &  0.6 & 18.2 & 4.2 & 6.5 & 1.5 & 7.2 & 6.3 \\
		& 10 & $\bar{\nsched}$ = 19    & 1.0 & 33.6 & 9.6 & 20.7 & 3.4 & 15.8 & 20.3 \\
		& 20 & &  1.2 & 49.1 & 16.0 & 61.4 & 6.2 & 31.9 & 60.5 \\
		& 50 &  & 1.4 & 52.8 & 17.7 & 64.9 & 7.7 & 37.8 & 65.0 \\
		
		\hline
	\end{tabular}
\end{table*}

We study the interplay between parallelism and the decline ratio of \algtop\ and other common placement algorithms.
We let \algtop\ adapt the number of schedulers according to its estimate of the system utilization and report the throughput of \algtop, captured by the average number of active schedulers that handle requests.
For the competing algorithms, we consider various values for the (fixed) number of schedulers.

Table~\ref{table:aspr_quality} summarizes the results. 
The algorithms DistFromDiag and Adaptive are abbreviated by Diag and Adapt, respectively.
The average number of active schedulers used by \algtop\ is indicated below its decline ratio.
Note that \algtop's decline ratio is always within the SLA requirement ($\dr =5\%$), and it uses between 14 and 20 active schedulers on average. 
Since the average number of arriving requests per cycle is $\lambda_a=20$,
it follows that it might be beneficial to occasionally have more than 20 schedulers to handle bursts of arrivals, but only 20 requests arrive per time unit on average. 
Random and \algtop\ yield the lowest decline ratio, both within the SLA constraint but
the communication overhead of \algtop\ is much lower than that of Random: the total number of queries made by {\em all} the schedulers which \algtop\ uses is the same as that of a {\em single} scheduler of Random. 
We note that this less accurate view of the system state causes \algtop's decline ratio sometimes to be slightly higher than that of Random (although always within the SLA).

\definecolor{LightCyan}{rgb}{0.88,1,1}
\begin{table*}[t!]
	\centering
    \caption{Total number of queries, throughput, and actual decline ratios of \algtop\ versus Random.}
	\TableFontSize
    \label{table:aspr_price}
	\renewcommand\arraystretch{1.3}
	\begin{tabular}{c|c|c|c||c|c|c|}
	\cline{2-7}
		&\multicolumn{3}{c||}{ \algtop} &\multicolumn{3}{c|}{Random} \\ \cline{2-7}
		&\multicolumn{3}{c||}{Target Decline Ratio ($\dr$)} &\multicolumn{3}{c|}{Number of Schedulers}\\
		 \cline{2-7}
		 
		 & 3\% & 5\% & 10\% & 1 &  10 & 20 \\ 
		  \cline{2-7}
	\rowcolor{LightCyan}	 &\multicolumn{6}{c|}{ NFV}\\
		 \cline{2-7}

		\hline			\multicolumn{1}{|c|}{Number of Queries} & 1553K  & 811K & 578K & \multicolumn{3}{c|}{11000K}\\
		\hline
		
			\multicolumn{1}{|c|}{Throughput [req./slot]} & 7.2 & 14 & 19.6 & 1  & 10 & 19.8 \\
		\hline

		\multicolumn{1}{|c|}{Decline Ratio ($\delta$)} & 0.4\%  & 0.4\% & 0.6\% & 0.3\% & 0.5\%& 0.8\%\\
		\hline

	\rowcolor{LightCyan}	 &\multicolumn{6}{c|}{ Google}\\
		\hline			\multicolumn{1}{|c|}{Number of Queries} & 3920K  & 3860K & 3823K & \multicolumn{3}{c|}{74724K}\\
			\hline
			\multicolumn{1}{|c|}{Throughput [req./slot]} & 19.8 & 19.9 & 19.9 & 1  & 10 & 19.9 \\
		\hline
		\multicolumn{1}{|c|}{Decline Ratio ($\delta$)} & 3.0\%  & 3.1\% & 2.9\% & 2.3\% & 2.4\%& 2.4\%\\
		\hline

		 \rowcolor{LightCyan}
		 & \multicolumn{6}{c|}{ Amazon}\\
		\hline			\multicolumn{1}{|c|}{Number of Queries} & 469K  & 370K & 354K & \multicolumn{3}{c|}{6745K}\\
			\hline
			\multicolumn{1}{|c|}{Throughput [req./slot]} & 15.3 & 19.3 & 19.9 & 1  & 10 & 19.9 \\
		\hline

		\multicolumn{1}{|c|}{Decline Ratio ($\delta$)} & 0.7\%  & 0.8\% & 1.0\% & 0.5\% & 1.0\% & 1.4\%\\
		\hline
	\end{tabular}
\end{table*}

\normalsize
Table~\ref{table:aspr_price} compares the throughput, the decline ratios, and the total number of queries of \algtop\ and Random.
%Like all other existing placement algorithms, Random queries all the hosts for every request. In contrast, \algtop\ queries at most $\left \lfloor \frac{\nbins}{\nsched} \right \rfloor $ hosts per request. 
Note that \algtop\ reduces the total number of queries by at least 85\%. Increasing \algtop's target decline ratio increases its parallelism, which in turn increases the throughput. This tradeoff highlights the tension between the decline ratio and the degree of parallelism. 
%In practice, it also increases the empirical decline ratio as more schedulers select the same host in parallel.  
% increases the level of parallelism and throughput 
% Furthermore, \algtop\aims at maximizing the number of schedulers (while maintaining the target decline ratio guarantee), so as to boost throughput. 
The best achievable throughput is 20, as it is the average arrival rate.  Indeed, \algtop\ and Random with fixed 20 schedulers are very close to the maximal throughput. Also, recall that, unlike Random, \algtop\ may fail due to not finding an available host in the queried hosts; thus, its decline ratio is sometimes higher. % is sometimes slightly worse than Random. %Thus, in the 
\newcommand {\subfloatwidth}  {0.72*\columnwidth}
\newcommand {\subfloatheight} {2 cm}
\newcommand {\marksize}       {1 pt}

\insertFullSection{
    \subsection{Under the hood of \algtop}
    Our next set of experiments studies the interplay between the system's utilization and the level of parallelism offered by \algtop. For these experiments, we use solely the NFV dataset. 
    
    Fig.~\ref{fig:algtop_paral_Vs_Util_inifinite_LT_Kmin} depicts the number of schedulers and the system utilization of \algtop. Initially, \algtop\ allows many schedulers as there are many available hosts for any flavor. As the utilization increases and the number of available hosts decreases, \algtop\ gradually reduces the number of schedulers. Intuitively, reducing the number of schedulers serves two goals: First, it allows each scheduler to query more hosts while still complying with the budget constraint. This increases the probability of finding an available host.  Second, having fewer schedulers reduces the collision probability. 
    
    \input{figs/tikz_sim_inifinite_LT}

    Recall that \algtop\ uses a conservative approach in estimating the number of available hosts ($\nfreebins$). This conservative approach indeed yields a very low decline ratio ($0.4\%$) -- but at the cost of throttling parallelism when utilization ramps up.
   We, therefore, consider a variant of \algtop, which we dub \algtopKavg. As its name suggests, this variant differs from Algorithm~\ref{alg:estimateK} in Line~\ref{alg:algtop:assign_k_n}, where it updates $\nfreebins$ according to the {\em average} number of available hosts taken over all flavors.

    Fig.~\ref{fig:algtop_paral_Vs_Util_inifinite_LT_Kavg} shows that \algtopKavg\ allows a significantly higher number of schedulers than \algtop, for any given level of utilization. As a result, \algtopKavg\ finishes handling all requests much faster than \algtop, implying a higher throughput. Indeed, switching from \algtop\ to \algtopKavg\ doubles the actual decline ratio to $0.8\%$ -- but this value is still well below the target decline ratio of $\dr = 5\%$. 
    
    \input{figs/tikz_sim_finite_LT}

    Our next experiment explores how both \algtop\ and \algtopKavg\ dynamically adjust the number of schedulers when the utilization fluctuates. To generate fluctuations in the utilization, we modeled the request arrival process as a variant of a \emph{Markov Modulated Poisson Process (MMPP)}~\cite{mmpp}. Specifically, the number of requests arriving per slot is a Poisson process with mean $\lambda_a$ throughout the experiment. However, for the first $20\%$ of the requests we use $\lambda_a=20$, to fill up the system; while for the rest of the requests we fix $\lambda_a=5$. Furthermore, in this experiment, requests have a finite lifetime. That is, the number of allocated requests leaving per time slot follows a Poisson process with mean $\lambda_d = 4$. Finally, we use here 100 replicas of the NFV dataset 
    (instead of 30 used in the rest of this section) so that even when requests leave their hosts, the hosts become utilized again with more arriving requests. These settings are intended to let utilization first build-up, and then stay at some (high) level, with mild fluctuations. 
    The results of this experiment are shown in Fig.~\ref{fig:algtop_paral_Vs_Util_finite_LT}. Both \algtop\ (Fig.~\ref{fig:algtop_paral_Vs_Util_finite_LT_Kmin}) and \algtopKavg\
    (Fig.~\ref{fig:algtop_paral_Vs_Util_finite_LT_Kavg})
    dynamically adapt the number of allowed schedulers to the utilization. However, \algtopKavg\ allows more schedulers than \algtop. It obtains shorter total run-time but experiences a higher decline ratio ($1.6\%$ for \algtopKavg\ versus $0.01\%$ for \algtop). Note that both algorithms are below the maximum allowed decline ratio ($5\%$).
     
    %\subsection{Effect of the Query Budget on Parallelism}
    We now investigate the effect of the query budget $\budget$ on the number of schedulers.
    We use the same settings of long-lived VMs as in the experiment used for the results presented in Table~\ref{table:aspr_quality}.
    In this experiment, we vary the budget $B$ on the overall number of accesses made by all the schedulers from $20\%$ to $100\%$ of the number of hosts.
    We report the level of parallelism, captured by the average number of active schedulers which \algtop\ employs along the run. Table~\ref{table:aspr_budget} illustrates the results. Indeed, parallelism is proportional to the given budget. However, the benefit of having a larger budget exhibits the effect of diminishing returns.
    Still, APSR runs more than ten schedulers, with an overhead that is 50\% smaller than that incurred by a {\em single} OpenStack scheduler.

    \begin{table}[ht]
        \caption{\algtop's throughput per budget \textit{B}, given as a percentage of total number of hosts ($\dr = 5\%$).} 
        %while the decline ratio is under $0.1\%$ in all configurations.}
        \label{table:aspr_budget}
    	\centering
    	\small
    	\begin{tabular}{c|c|c|c|c|c|}
    	\cline{2-6}
    		&\multicolumn{5}{c|}{ Budget } \\
    		 \cline{2-6}
    		 & 20\% & 40\% & 60\% & 80\% &  100\% \\ 
    		 	\hline
    			\multicolumn{1}{|c|}{Throughput [req./slot]} & 
    			6.5 & 9.8 & 11.7 & 12.8 &  14\\
    
    		\hline
    	\end{tabular}
    	\vspace{0.2cm} % increase space after table to ensure it doesn't "stick" to following text
    \end{table}
    
}

\insertShortSection{
    \textbf{OpenStack Evaluation: }
    \label{sec:OpenStack}
    We now evaluate \algtop\ in an OpenStack environment (Mitaka release)~\cite{openstack} on an HP ProLiant BL460c Gen9 server with two Intel(R) Xeon(R) E5-2680v4 processors with 28 cores (56 cores total) running at 2.4 GHz, and a total RAM of 256GB.
    We run a functional scheduler implementation and use OpenStack's Benchmarking to emulate the remote hosts~\cite{nova_emulator}. 
    We periodically send 200 request batches from the NFV dataset and wait for the scheduler to place all of them. In total, we send $2800$ requests to place on 400 hosts, attaining a resource utilization of $\approx$90\% at a steady state.  
    We set \algtop's parameters to $T=10 sec$, $\budget = 100$ and $\dr \in \{2\%, 3\%, 5\%\}$. 
    
    Table~\ref{table:aspr_query_retry} compares the throughput, decline ratio, and the total number of queries of \algtop\ and the default Filter scheduler. 
    The table shows that \algtop's decline ratio is always within the target bound. Furthermore, \algtop\ attains a similar throughput to running 8 Filter schedulers in parallel while keeping a much lower decline ratio than that presented by 8 Filter schedulers. Finally, \algtop\ reduces the number of host queries by $\approx 90\%$.  
    
}

    \begin{table*}[t!]
    \centering
        \caption{OpenStack: Number of queries, average number of schedulers, throughput, and actual decline ratios of \algtop\ and Filter Scheduler for the NFV dataset.}
        \label{table:aspr_query_retry}
    \renewcommand\arraystretch{1.3}
    \TableFontSize
    \begin{tabular}{c|c|c|c||c|c|c|}
        \cline{2-7}
    	&\multicolumn{3}{c||}{ \algtop} &\multicolumn{3}{c|}{Filter Scheduler} \\ \cline{2-7}
    	&\multicolumn{3}{c||}{Target decline ratio ($\dr$)} &\multicolumn{3}{c|}{Number of schedulers}\\ \cline{2-7}
    	 
    	 & 2\% & 3\% & 5\% & 1 & 8 & 16 \\ 
    
    	\hline
    	\multicolumn{1}{|c|}{Number of queries} & 110K & 102K & 108K & \multicolumn{3}{c|}{2240K}\\
    	\hline
    	\multicolumn{1}{|c|}{Avg \# of schedulers allowed} & 13.5 & 16 & 16 & 1 & 8 & 16\\
    	\hline
    	\multicolumn{1}{|c|}{Throughput [req./sec.]} & 2.6 & 2.8 & 2.6 & 1 & 2.6 & 2.7\\
    	\hline
        \multicolumn{1}{|c|}{Actual decline ratio ($\delta$)} & 1.0\% & 0.7\% & 0.7\% & 0\% & 3.8\%& 13.6\%\\
    	\hline
    \end{tabular}
    \end{table*}

\insertFullSection{
    \subsection{OpenStack Evaluation}\label{sec:OpenStack}
    We now evaluate \algtop\ in an OpenStack environment (Mitaka release)~\cite{openstack} on an HP ProLiant BL460c Gen9 server with two Intel(R) Xeon(R) E5-2680v4 processors with 28 cores (56 cores total) running at 2.4 GHz, and a total RAM of 256GB.
    We run a functional scheduler implementation and use OpenStack's Benchmarking to emulate the remote hosts~\cite{nova_emulator}. 
    We periodically send 200 request batches from the NFV dataset and wait for the scheduler to place all of them. 
    We set \algtop's parameters to $T=10 sec$, $\budget = 100$ and $\dr \in \{2\%, 3\%, 5\%\}$.

    Table~\ref{table:aspr_query_retry} compares the throughput, decline ratio, and the total number of queries of \algtop\ and the default Filter scheduler. 
    The table shows that \algtop's decline ratio is always within the target bound. Furthermore, \algtop\ attains a similar throughput to running 8 Filter schedulers in parallel while keeping a much lower decline ratio than that presented by 8 Filter schedulers. Finally, \algtop\ reduces the number of host queries by $\approx 90\%$.  

    \ignore{
        Figure~\ref{fig:under_the_hood_sched} shows the number of schedulers used during the experiment for different target decline ratios ($\dr \in \{2\%, 3\%\}$). As can be observed, for $\dr =3\%$ the number of schedulers is always $16$, which implies that \algtop\ can utilize more than $16$ schedulers. For $\dr= 2\%$, \algtop\ retains more than $10$ schedulers throughout the experiment.
    }
    
    \ignore{    
        \Figure[ht]
            [width=\columnwidth]
            {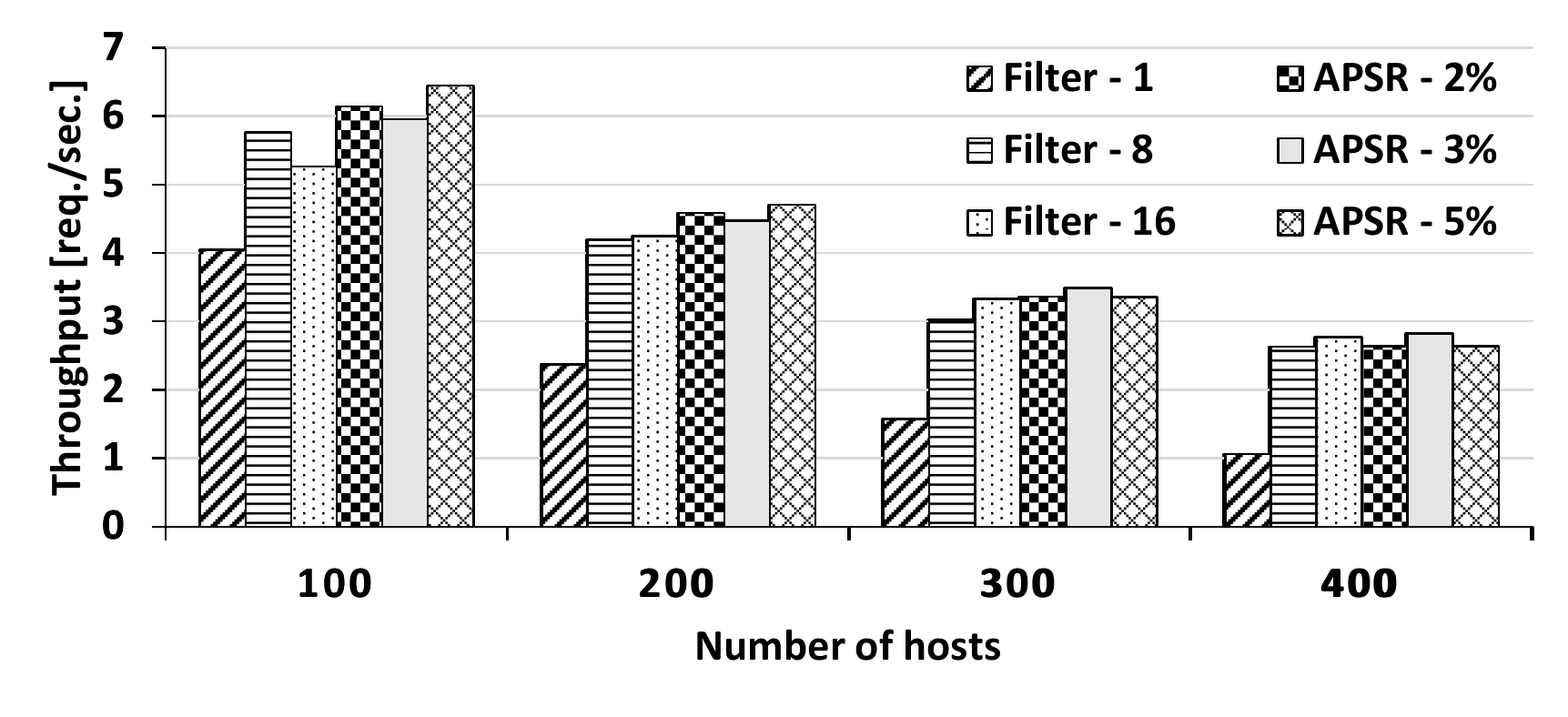}
            {OpenStack: throughput while varying the number of hosts for the NFV dataset.
            \label{fig:OS_throughput}
        }
    
        \Figure[ht]
            [width=\columnwidth]
            {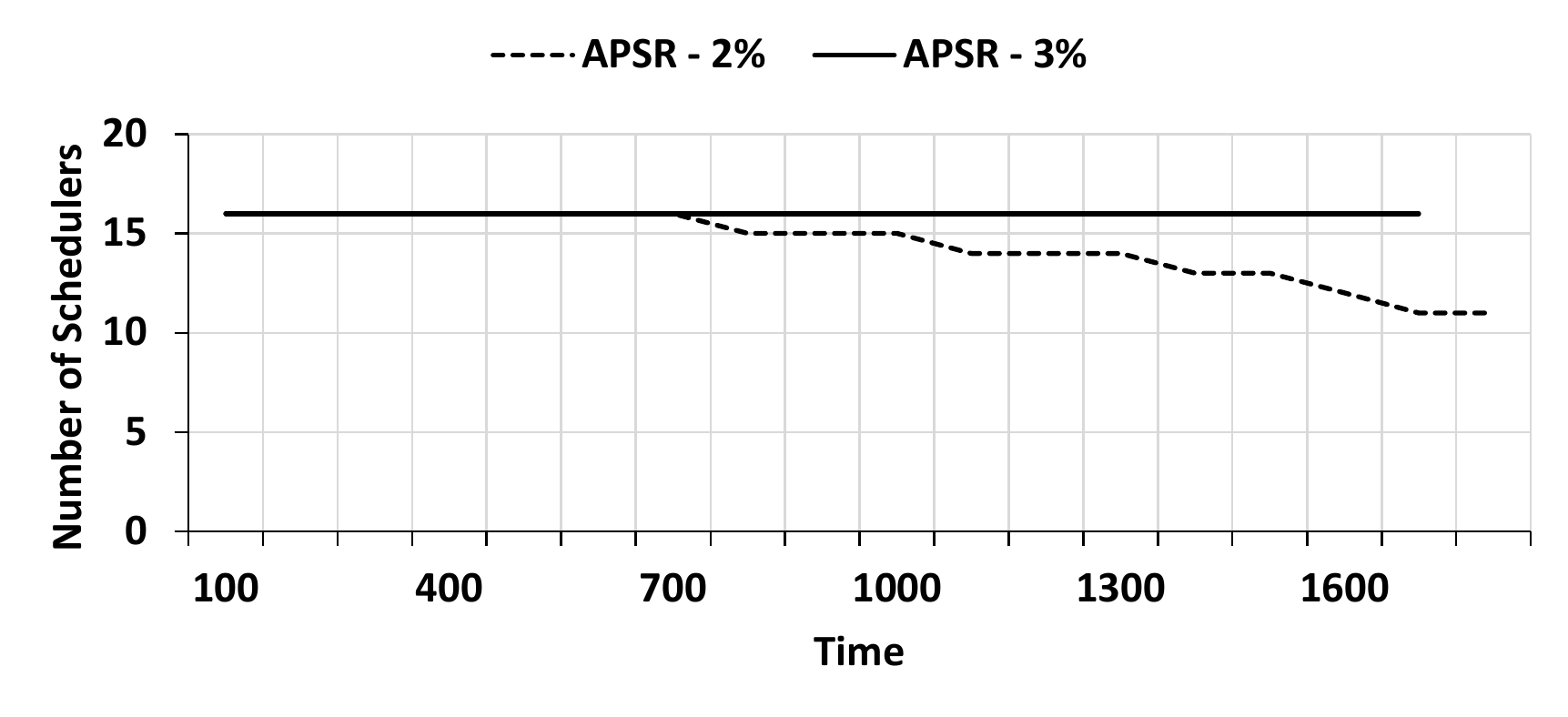}
            {OpenStack: number of schedulers over time while varying the target decline ratio ($\dr$) for the NFV dataset.
            \label{fig:under_the_hood_sched}
        }
    }
    
 Our experiments suggest that in current OS implementation, solving the bottleneck of parallelism addressed in our work,
    raises new challenges and bottlenecks. Therefore, to fully exploit the benefits of our work, a further investigation of the bottlenecks in OS is required. 
    
}

\section {Discussion, Conclusions and Future Work}
\label{sec:conclusions}
Our work seeks high-throughput placement of virtual machines to better cope with long service chains.
Parallelism improves throughput, but many placement algorithms behave poorly in parallel settings. 
Our \algtop\ algorithm implements random placement while minimizing the communication overhead and dynamically adjusting the degree of parallelism to ensure that decline ratios satisfy their SLA requirements. 
We formally prove the correctness of \algtop\ and provide insights into the possibilities and limitations of parallel resource management. 

We evaluate \algtop\ on three real workloads and demonstrate its capability to provide high degrees of parallelism with small decline ratios and low communication overheads. 
We then integrate \algtop\ into the OpenStack cloud management platform. 
We show that \algtop\ matches the best throughput of OpenStack's default Filter Scheduler while reducing the decline ratio from up to $13.6\%$ to $\approx 1\%$, and the communication overheads by $\approx 20$x. That is, \algtop\ also implies less clutter and drain on the system. 

Looking into the future, we observe that OpenStack only gains up to $3$x speedup from parallelism, whereas \algtop\ easily supports many parallel schedulers. Thus, we plan to carefully benchmark OpenStack, identify its current bottlenecks, and unleash its full potential for parallel resource management. 

\bibliographystyle{abbrv}
\bibliography{refs}

\end{document}

%% file: figs/tikz_sim_inifinite_LT.tex
\begin{figure}[ht]
    \subfloat [\label{fig:algtop_paral_Vs_Util_inifinite_LT_Kmin}\algtop\ (actual decline ratio is 0.4\%).] { % Fig.4a in NSDI 
    \begin{tikzpicture}
        \pgfplotsset{
            scale only axis,
            xmin=0, xmax=1000
        }
        
        \begin{axis}[
            width  = \subfloatwidth,
            height = \subfloatheight,
            axis y line*=left,
            ymin=0, ymax=1,
            xlabel = Time,
            ylabel = Utilization,
            legend     style = {font=\scriptsize},
            label      style = {font=\scriptsize},
            tick label style = {font=\scriptsize},
            ]
            \addplot[smooth,mark=triangle,mark size = 1.5pt] coordinates{
                (0, 0.00)
                % (10, 0.02)
                (20, 0.04)
                % (30, 0.05)
                (40, 0.07)
                % (50, 0.08)
                (60, 0.10)
                % (70, 0.11)
                (80, 0.13)
                % (90, 0.14)
                (100, 0.16)
                % (110, 0.17)
                (120, 0.19)
                % (130, 0.20)
                (140, 0.22)
                % (150, 0.24)
                (160, 0.25)
                % (170, 0.27)
                (180, 0.28)
                % (190, 0.30)
                (200, 0.31)
                % (210, 0.33)
                (220, 0.34)
                % (230, 0.36)
                (240, 0.38)
                % (250, 0.39)
                (260, 0.41)
                % (270, 0.42)
                (280, 0.44)
                % (290, 0.45)
                (300, 0.47)
                % (310, 0.48)
                (320, 0.49)
                % (330, 0.51)
                (340, 0.53)
                % (350, 0.54)
                (360, 0.55)
                % (370, 0.56)
                (380, 0.57)
                % (390, 0.59)
                (400, 0.60)
                % (410, 0.61)
                (420, 0.63)
                % (430, 0.64)
                (440, 0.66)
                % (450, 0.67)
                (460, 0.68)
                % (470, 0.70)
                (480, 0.72)
                % (490, 0.73)
                (500, 0.75)
                % (510, 0.76)
                (520, 0.77)
                % (530, 0.79)
                (540, 0.80)
                % (550, 0.82)
                (560, 0.84)
                % (570, 0.85)
                (580, 0.86)
                % (590, 0.88)
                (600, 0.89)
                % (610, 0.89)
                (620, 0.90)
                % (630, 0.91)
                (640, 0.91)
                % (650, 0.92)
                (660, 0.92)
                % (670, 0.92)
                (680, 0.93)
                % (690, 0.93)
                (700, 0.94)
                % (710, 0.94)
                (720, 0.94)
                % (730, 0.94)
                (740, 0.94)
                % (750, 0.95)
                (760, 0.95)
                % (770, 0.95)
                (780, 0.95)
                % (790, 0.95)
                (800, 0.95)
                % (810, 0.95)
                (820, 0.95)
                % (830, 0.95)
                (840, 0.95)
                % (850, 0.95)
                (860, 0.96)
                % (870, 0.96)
                (880, 0.96)
                % (890, 0.96)
                (900, 0.96)
                % (910, 0.96)
                (920, 0.96)
                % (930, 0.96)
                }; \label{plot_util1}
        \end{axis}
        
        \begin{axis}[legend columns=1,
            width  = \subfloatwidth,
            height = \subfloatheight,
            axis y line*=right,
            axis x line=none,
            ymin=0, ymax=100,
            ylabel style={align=center}, 
            ylabel = \# of schedulers \\allowed by \algtop,
            ytick={20, 40, 60, 80, 100},
            legend style     = {font=\scriptsize, at={(0.78,0.65)},anchor=north,legend columns=-1},
            label style      = {font=\scriptsize},
            tick label style = {font=\scriptsize},
            ]
            
        \addlegendimage{/pgfplots/refstyle=plot_util1}\addlegendentry{Utilization}
            \addplot[smooth,mark=*,mark size = 1.5pt] coordinates{
                (0, 87)
(20, 87)
(40, 86)
(60, 85)
(80, 84)
(100, 83)
(120, 82)
(140, 80)
(160, 78)
(180, 76)
(200, 74)
(220, 71)
(240, 69)
(260, 66)
(280, 63)
(300, 60)
(320, 57)
(340, 54)
(360, 51)
(380, 48)
(400, 45)
(420, 42)
(440, 39)
(460, 36)
(480, 33)
(500, 29)
(520, 26)
(540, 23)
(560, 20)
(580, 16)
(600, 14)
(620, 11)
(640, 9)
(660, 7)
(680, 6)
(700, 5)
(720, 4)
(740, 3)
(760, 2)
(780, 2)
(800, 2)
(820, 1)
(840, 1)
(860, 1)
(880, 1)
(900, 1)
(920, 1)
            }; \addlegendentry {\# Schedulers}
        \end{axis}
    \end{tikzpicture}
    } % End of subfloat
%%%%%%%%%%%%%%%%%%%%%%%%%%%%%%%%%%%%%%%%%%%%%%%%%%%%%%%%%%%%%%%%%%%%%%%%%%%%%%%%%%%%%%%%%%%%%%%%%%%%%%%%%%%%%%%%%%%%%%%%%%%%%%%%%%%%%%%%%%%

    \subfloat [\label{fig:algtop_paral_Vs_Util_inifinite_LT_Kavg}\algtopKavg\ (actual decline ratio is 0.8\%).] {% Fig.4b in NSDI
    \begin{tikzpicture}
        \pgfplotsset{
            scale only axis,
            xmin=0, xmax=1000
        }
        
        % APSR  (actual decline ratio is 0.4\%).
        \begin{axis}[
            width  = \subfloatwidth,
            height = \subfloatheight,
            axis y line*=left,
            ymin=0, ymax=1,
            xlabel = Time,
            ylabel = Utilization,
            legend     style = {font=\scriptsize},
            label      style = {font=\scriptsize},
            tick label style = {font=\scriptsize},
            ]
            \addplot[smooth,mark=triangle,mark size = 1.5pt] coordinates{
                (0, 0.00)
                % (10, 0.02)
                (20, 0.04)
                % (30, 0.05)
                (40, 0.07)
                % (50, 0.08)
                (60, 0.10)
                % (70, 0.11)
                (80, 0.13)
                % (90, 0.14)
                (100, 0.16)
                % (110, 0.17)
                (120, 0.19)
                % (130, 0.20)
                (140, 0.22)
                % (150, 0.24)
                (160, 0.25)
                % (170, 0.27)
                (180, 0.28)
                % (190, 0.30)
                (200, 0.31)
                % (210, 0.33)
                (220, 0.34)
                % (230, 0.36)
                (240, 0.38)
                % (250, 0.39)
                (260, 0.41)
                % (270, 0.42)
                (280, 0.43)
                % (290, 0.45)
                (300, 0.47)
                % (310, 0.48)
                (320, 0.49)
                % (330, 0.51)
                (340, 0.52)
                % (350, 0.54)
                (360, 0.55)
                % (370, 0.56)
                (380, 0.57)
                % (390, 0.59)
                (400, 0.60)
                % (410, 0.61)
                (420, 0.63)
                % (430, 0.64)
                (440, 0.66)
                % (450, 0.67)
                (460, 0.68)
                % (470, 0.70)
                (480, 0.71)
                % (490, 0.73)
                (500, 0.74)
                % (510, 0.76)
                (520, 0.77)
                % (530, 0.79)
                (540, 0.80)
                % (550, 0.81)
                (560, 0.83)
                % (570, 0.84)
                (580, 0.86)
                % (590, 0.87)
                (600, 0.88)
                % (610, 0.89)
                (620, 0.90)
                % (630, 0.91)
                (640, 0.92)
                % (650, 0.93)
                (660, 0.94)
                }; \label{plot_util2}
        \end{axis}
        
        \begin{axis}[
            width  = \subfloatwidth,
            height = \subfloatheight,
            axis y line*=right,
            axis x line=none,
            ymin=0, ymax=100,
            ylabel style={align=center}, 
            ylabel = \# of schedulers \\allowed by \algtop,
            ytick={20, 40, 60, 80, 100},
            legend style     = {font=\scriptsize, at={(0.6,0.3)},anchor=north,legend columns=-1},
            label style      = {font=\scriptsize},
            tick label style = {font=\scriptsize},
            ]
            
        \addlegendimage{/pgfplots/refstyle=plot_util2}\addlegendentry{Utilization}
            \addplot[smooth,mark=*,mark size = 1.5pt] coordinates{
                (0, 87)
                % (10, 87)
                (20, 87)
                % (30, 87)
                (40, 87)
                % (50, 87)
                (60, 86)
                % (70, 86)
                (80, 86)
                % (90, 86)
                (100, 86)
                % (110, 86)
                (120, 86)
                % (130, 86)
                (140, 85)
                % (150, 85)
                (160, 85)
                % (170, 85)
                (180, 84)
                % (190, 84)
                (200, 84)
                % (210, 84)
                (220, 83)
                % (230, 83)
                (240, 83)
                % (250, 82)
                (260, 82)
                % (270, 81)
                (280, 81)
                % (290, 80)
                (300, 80)
                % (310, 79)
                (320, 79)
                % (330, 78)
                (340, 77)
                % (350, 77)
                (360, 76)
                % (370, 75)
                (380, 75)
                % (390, 74)
                (400, 73)
                % (410, 72)
                (420, 72)
                % (430, 71)
                (440, 70)
                % (450, 69)
                (460, 68)
                % (470, 67)
                (480, 66)
                % (490, 65)
                (500, 64)
                % (510, 63)
                (520, 62)
                % (530, 60)
                (540, 59)
                % (550, 58)
                (560, 56)
                % (570, 55)
                (580, 53)
                % (590, 52)
                (600, 50)
                % (610, 48)
                (620, 47)
                % (630, 45)
                (640, 43)
                % (650, 41)
                (660, 40)
            }; \addlegendentry {\# Schedulers}
        \end{axis}
    \end{tikzpicture}
    } % End of subfloat
	\caption{Cloud resource utilization and the number of schedulers in APSR for the NFV dataset under Poisson arrivals (requests have infinite lifetime).}
	\label{fig:algtop_paral_Vs_Util_infinite_LT}
\end{figure}
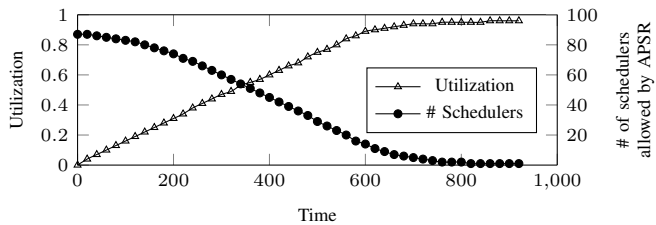
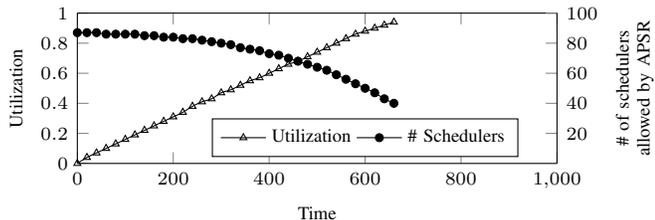